# Algorithmic modelling of a complex redundant multi-state system subject to multiple events, preventive maintenance, loss of units and a multiple vacation policy through a MMAP

Juan Eloy Ruiz-Castro[a] and Hugo Alaín Zapata-Ceballos[b]

[a] Dept. of Statistics and Operations Research and IMAG, University of Granada, Spain. e-mail:jeloy@ugr.es

[b] Universidad de Sucre, Colombia. e-mail: hugo.zapata@unisucre.edu.co

## Abstract

A complex multi-state redundant system undergoing preventive maintenance and experiencing multiple events is being considered in a continuous time frame. The online unit is susceptible to various types of failures, both internal and external in nature, with multiple degradation levels present, both internally and externally. Random inspections are continuously monitoring these degradation levels, and if they reach a critical state, the unit is directed to a repair facility for preventive maintenance.

The maintenance place is managed by a repairperson, who follows a multiple vacation policy dependent on the operational status of the units. The repairperson is responsible for two primary tasks: corrective repair and preventive maintenance. The time durations within the system follow phase-type distributions, and the model is constructed utilizing Markovian Arrival Processes with marked arrivals. A variety of performance measures, including transient and stationary distributions, are calculated using matrix-analytic methods. This methodology allows for the representation of significant outcomes and the general behavior of the system in a matrix-algorithmic structure.

To enhance the model's efficiency, both costs and rewards are incorporated into the analysis. A numerical example is presented to showcase the model's flexibility and effectiveness in real-world applications.

ACRONYMS

| | | |
|---|---|---|
| $(\boldsymbol{\alpha}, \mathbf{T})$ | : | Phase type distribution for the internal performance with order $m$ |
| $(\boldsymbol{\gamma}, \mathbf{L})$ | : | Phase type distribution for the time between two consecutive external shocks (order $t$) |
| $\omega^0$ | : | Probability of direct total failure after an external shock |
| $\mathbf{W}$ | : | Transition probability matrix for any two internal states triggered by an external shock |
| $\mathbf{D}$ | : | Transition probability matrix for the cumulative damage after an external shock (order *d*) |
| $(\eta,\mathbf{M})$ | : | Phase type distribution for the time between two consecutive inspections (order *m*) |
| $(\upsilon,\mathbf{V})$ | : | Phase type distribution for the duration of vacation time (order *v*) |
| $(\boldsymbol{\beta}^1,\mathbf{S}_1)$ | : | Phase type distribution for the corrective repair time (order $z_1$) |
| $(\boldsymbol{\beta}^2,\mathbf{S}_2)$ | : | Phase type distribution for the preventive maintenance time (order $z_2$) |
| $\mathbf{D}^Y$ | : | Generator block corresponding to the embedded Markov chain within the **MMAP** for the event *Y* |
| $\mathbf{D}_k^Y$ | : | Generator matrix block for the transitions between macro-states *k* units for the event *Y* |
| $\mathbf{D}_{ij}^{Y,k,v(nv)}$ | : | Generator matrix block for the transitions between macro-states, *k* units within the system, from *i* units in the repair facility to *j*, when the repairperson is on vacation (*v*) or not (*nv*), when the event *Y* is produced. |
| $\mathbf{D}$ | : | Generator matrix for the embedded Markov chain |
| $\phi$ | : | Initial distribution for the system |
| $\mathbf{p}_{\mathbf{E}_s^{k,x}}(t)$ | : | Transient distribution for the macro-state *k* units in the system, *s* of them in the repair facility and the repairperson is on vacation (*x*=*v*) or not (*x*=*nv*). |
| $\boldsymbol{\pi}_{\mathbf{E}_s^{k,x}}$ | : | Stationary distribution for the macro-state *k* units in the system, *s* of them in the repair facility and the repairperson is on vacation (*x*=*v*) or not (*x*=*nv*). |
| $A(t)$ | : | Availability at time *t* |
| $A$ | : | Availability in stationary regime |
| $\Psi_{k,s}^x(t)$ | : | Mean time that the system is with *k* units in the system, *s* of them in the repair facility, with the repairperson on vacation or not (*x*=*v* or *x*=*nv*, respectively) up to time *t* |
| $\Psi_{k,s}^x$ | : | Proportional time that the system is with *k* units in the system, *s* of them in the repair facility, with the repairperson on vacation or not (*x*=*v* or *x*=*nv*, respectively) in stationary regime |
| $\Psi_{k,s}(t)$ | : | Mean time that the system is with *k* units in the system, *s* of them in the repair facility, up to time *t* |
| $\Psi_{k,s}$ | : | Proportional time that the system is with *k* units in the system, *s* of them in the repair facility, in stationary regime |
| $\Psi_k(t)$ | : | Mean time that the system is with *k* units in the system up to time *t* |
| $\Psi_k$ | : | Proportional time that the system is with *k* units in the system in stationary regime |
| $\mu_{op}(t)$ | : | Mean operational time up to time *t* |
| $\Lambda^{rep}(t)$ ; $\Lambda^{rep}$ | : | Mean number of repairable failures in transient and stationary regime |
| $\Lambda^{mi}(t)$ ; $\Lambda^{mi}$ | : | Mean number of major inspections in transient and stationary regime |
| $\Lambda^{nr}(t)$ ; $\Lambda^{nr}$ | : | Mean number of non-repairable failures in transient and stationary regime |
| $\Lambda^{ret}(t)$ ; $\Lambda^{ret}$ | : | Mean number of returns to work in transient and stationary regime |
| $\Lambda^{ret-be}(t)$ ; $\Lambda^{ret-be}$ | : | Mean number of returns and start a new vacation period in transient and stationary regime |
| $\Lambda^{after}(t)$ ; $\Lambda^{after}$ | : | Mean number of vacation periods after repair in transient and stationary regime |
| $\Lambda^{NS}(t)$ ; $\Lambda^{NS}$ | : | Mean number of new systems in transient and stationary regime |
| **nr** | : | Net reward vector (only costs and rewards from the online unit) |
| **nc** | : | Cost vector according to the states of the system |
| **c** | : | Net reward/cost vector according to the states of the system |
| $\Phi_w(t)$ ; $\Phi_w$ | : | Mean net profit up to time *t* and rate per unit of time in stationary regime |
| $\Phi_{rf}(t)$ ; $\Phi_{rf}$ | : | Mean cost up to time *t* and rate per unit of time in stationary regime |
| $\Phi(t)$ ; $\Phi$ | : | Mean net total profit up to time *t* and rate per unit of time in stationary regime |

# 1. Introduction

In the field of reliability, it is of interest to incorporate different maintenance structures in order to prevent repairable or non-repairable failures that may cause personal injury and/or significant costs.

This fact is also relevant for multi-state systems. The integration of multi-state systems into reliability, extending beyond the binary scenario, is now a recognized practice. A multi-state system (MSS) refers to a system capable of operating at multiple levels and experiencing various failure modes, each with distinct impacts on the system's overall performance. Lisnianski et al. (2010) presented a comprehensive MSS reliability theory, built upon accomplishments in this field, and included a variety of significant case studies. Recently, Wu et al. (2024) investigated a reliability model and the optimal preventive maintenance policy for a multi-state performance-sharing system subject to random shocks. Additionally, Levitin et al. (2024) introduced a new numerical algorithm aimed at evaluating the mission success probability of the analyzed multistate production-storage system under the corrective maintenance policy. This algorithm was implemented to optimize the corrective maintenance policy, with the goal of maximizing the mission success probability.

To improve the reliability of a general system, particularly a multi-state system, it is common practice to use redundant systems and implement multiple preventive maintenance protocols for them. There are multiple redundant systems that can be incorporated to extend the reliability of a system. Models of systems with units in hot, cold, and warm standby, as well as $k$-out-of-$n$: $G$ systems can be found in the literature. Recently, Guo et al. (2023) consider the suspended animation rule in a series model composed of Markovian k-out-of-N: G warm standby subsystems. Also, Gao (2023) studies and optimizes the behavior of a system with dependent failures, units in cold and warm standby, and two types of repairers, regular and expert, using Markov renewal theory. Also, Levitin et al. (2023) model a 1-out-of-N standby system with resource-constrained elements experiencing prescheduled mode transfers. For this purpose, a numerical algorithm is proposed to optimize the mode transfer problem by minimizing the expected downtime of the redundant system.

On the other hand, preventive maintenance acts as a proactive shield against unexpected equipment breakdowns and failures, reducing costs for both repairable and non-repairable systems. Therefore, a well-designed preventive maintenance policy enhances system efficiency, minimizes downtime, improves reliability, extends equipment lifespan, and enhances safety. Maintenance policies for reliability systems have been extensively discussed in Nakagawa (2005), providing thorough insights into the topic. Levitin et al. (2021) proposed a model for the transfer of time-consuming tasks in standby systems' event transition-based reliability analysis. In this method, preventive replacements are executed based on an optimized predetermined schedule to maximize reliability. In a different context, Yang et al. (2019) examined a strategy for pre-emptive maintenance targeting a solitary unit susceptible to malfunction arising from either internal decay or abrupt impact. Their study employed a non-uniform Poisson process, segmenting the internal failure progression into two phases. Cha et al. (2021) also considered various replacement policies in a system with worse-than-minimal repair. Such repair occurs in practice due to previous faulty repairs. This work incorporates the generalized Polya process of repairs for modelling.

Adhering to a strategic maintenance policy aimed at minimizing costs and downtime, it can be more beneficial for the system to incorporate periods of unavailability for the repairperson. A random vacation period is defined as the duration during which the repairperson is unavailable in a repair channel. One of the main challenges is optimizing the vacation time, as an excessively long period can escalate the system's non-operational time, leading to increased costs and reduced productivity. A well-designed vacation policy for the repairperson effectively balances the system according to the optimal distribution. Various strategies are under consideration for the repairperson's vacation periods, including the N-policy, multiple and single vacation policies, Bernoulli vacation policy, multiple and single working vacation policies, and vacation interruption policy, among others. Each of these policies adopts a distinct strategy in the modelling. It is crucial to acknowledge that these policies not only impact costs and rewards but also influence the reliability of the systems. Shekhar et al. (2020) showed comprehensive models and a thorough investigation of different vacation policies. Furthermore, Gao et al. (2023) analyzed the behaviour of a series system subject to common failures and introduced delayed vacation for the repairperson through Markov renewal processes. Utilizing the redundancy and repair facility features, Kumar et al. (2023) developed a double retrial orbit queuing model for the fault-tolerant machining system (FTMS) operating under the restriction of admission of repair jobs based on threshold policy and working vacation. Liu et al. (2015) studied a cold standby repairable system by considering Phase-type distributions and multiple vacation policy for the repairperson.

When modelling the behaviour of a complex multi-state system subject to multiple events, with redundant units, and incorporating preventive maintenance policies and downtime periods for the repairer, it is common for the probabilistic expressions obtained to be intractable and for the construction of measures and interpretation of results to be complex. A methodological option for studying such systems is to consider phase-type (PH) distributions and Markovian Arrival Processes (MAP). PH distributions were introduced by Neuts (1975, 1981) as the time until absorption in an absorbing Markov chain. This class of distributions has very good properties, and one particularly interesting property is that the class of PH distributions is dense in the set of non-negative probability distributions. This property allows for the consideration of general distributions. On the other hand, MAP processes were also introduced by Neuts (1979) as a generalization of counting processes. PH distributions and MAP processes enable analytic-matrix structures to facilitate well-structured modelling of complex reliability systems and subsequent interpretation of results. Ruiz-Castro and Dawabsha (2020) built redundant complex systems that evolved in discrete time by using MMAPS and PH distributions. Sophisticated complex systems incorporating preventive maintenance and multiple vacation policies, utilizing Marked Markovian Arrival Processes, have been developed. Ruiz-Castro (2019, 2022) included models for both discrete and continuous single unit systems, as well as for discrete multi-state redundant with loss of units (Ruiz-Castro, 2021). It is worth noting that recent research articles on the modeling of complex multi-state systems with preventive maintenance have become abundant, with Markov theory playing a significant role. Yang et al. (2021) presented an integrated optimization of production scheduling and preventive maintenance for single-machine multi-state systems using a reinforcement learning approach. The study formulated the problem as a Markov decision process and introduced a heuristic method to enhance efficiency. Two other recent works modeling $k$-out-of-$n$: $F$ systems are presented by Ning et al. (2024) and Dong et al. (2024). The first explores optimizing preventive maintenance and triggering mechanisms. The study introduced a Markov decision process to determine optimal inspection intervals and maintenance strategies, aiming to minimize costs and enhance system reliability. In the second, the paper investigates the reliability and preventive maintenance of a consecutive $k$-out-of-$n$: $F$ balanced multi-state system under shock environments. It introduces a model where system failure occurs when consecutive components fail due to external shocks. To minimize maintenance costs, the study proposes an optimization model for preventive maintenance. The model's effectiveness is demonstrated through finite Markov chain embedding and Phase-type distribution analysis.

This work models a real-life system with a main unit and an undetermined number of units in cold standby. It generalizes previous models developed in discrete time to continuous time, with modifications not being immediate. Additionally, new effects following external shocks are introduced, as well as new events in the MMAP, new performance measures in continuous time, and new costs that will allow for a better decision to optimize the model. The main unit can undergo multiple events such as repairable or non-repairable internal failures, external shocks resulting in total unit failure, modification of internal wear, or cumulative external damage. In the field of reliability, it is common to consider that when a non-repairable failure occurs, the unit is immediately replaced. In this work, this situation is not considered, the unit is not replaced as long as the system is operational. The level of degradation of the main unit, as well as the cumulative damage from external shocks, is partitioned into minor or major damage. These tiers are monitored through random inspections. If major damage is observed in any case, the unit is moved to the repair channel for preventive maintenance. Therefore, the repairer performs two distinct tasks: corrective repair and preventive maintenance. This complex system is optimized considering a multiple vacation policy for the repairer. For the development of the modelling and the construction of measures associated with the system, it has been considered that the implicit times in the system are phase-type distributed, and that the renewal process of the time between external shocks is also phase-type. To obtain measures regarding the quantification of events over time, the system has been modelled by constructing a Markov Arrival Process with marked arrivals (MMAP). The entire development is carried out algorithmically-matrix-wise, both in transient and stationary regimes. For the calculation of the steady-state distribution, analytical-matrix techniques have been considered, obtaining it algorithmically. The system is modelled and operational measures are constructed, and multiple costs/rewards are introduced in a vectorial form. To show the versatility of the model, an example is presented where the system is optimized according to vacation policy and preventive maintenance.

The rest of the work is distributed as follows. Section 2 presents the system with assumptions, describing the state space. Section 3 is focused on modelling by constructing the MMAP. Operational measures in transient and steady-state regimes are provided in Section 4. Section 5 introduces the costs and constructs associated measures. It is in Section 6 where the versatility of the model is shown by optimizing a system in a numerical example. Section 7 presents the conclusions.

## 2. The system

A complex multi-state system composed of $n$ units, the online one and the rest disposed in cold standby is assumed. The online unit has multiple states, including minor and major states depending on internal damage. A major state indicates a higher risk of failure.

The active unit is exposed to various events, such as internal failures, which can be repairable or non-repairable, as well as external shocks. External shocks can lead to different outcomes, including total failure, changes in internal performance, or repairable/non-repairable internal failures. Cumulative external damage is produced after an external shock crossing multiple external states. When a threshold is reached, a non-repairable failure occurs.

After a repairable failure, the unit is sent to the repair facility for corrective repair. The repair facility is composed of one repairperson, who may take vacations.

To prevent serious damage and significant economic losses, random inspections are conducted. Should the inspector detect significant damage while the unit remains active, it will be sent to the repair facility for preventive maintenance, which takes a certain amount of random time different to corrective repair. Therefore, the repairperson takes on two different tasks.

When an online unit fails, a cold standby unit replaces it. It begins without damage because, following repair or preventive maintenance, it's essentially brand new. The system experiences unit losses, and upon an irreparable failure, the unit is withdrawn. The system continues to function until all units are depleted. If only one unit remains and it undergoes an irreparable failure, the system is rebooted with $n$ units.

A vacation policy is introduced in the system to optimize rewards. The repairer stationed at the repair facility can undertake two tasks: corrective repairs and preventive maintenance. To enhance system efficiency, the repairer may take several vacations of varying durations, based on specific criteria.

Initially, all units are functioning, and the repair technician is on leave. Upon returning from vacation, a new vacation period begins if there are $R$ or more operational units in the system. Alternatively, if there are $k - R + 1 = N$ or more failed units requiring repair, where $k$ represents the total number of units in the system ($k = 1, ..., n$), the repair technician remains at the repair facility. After completing a repair, the repairer initiates a new vacation period if $R$ units are operational. Given the potential for unit losses in the system, the repairer must either remain at the repair facility or interrupt their vacation and return when the number of units in the system falls below $R$.

The next section outlines the system's assumptions.

### 2.1. Assumptions

**Assumption 1**. The time for internal performance of the online unit follows a PH distribution, represented as $(\boldsymbol{\alpha}, \mathbf{T})$, where $m$ represents the number of internal stages. The transition intensities between internal transient phases varies depending on the initial and ending phases. The column vectors $\mathbf{T}_r^0$ and $\mathbf{T}_{nr}^0$ contain the transition intensities from performance states to repairable and non-repairable failures, respectively.

**Assumption 2**. The internal performance of the online unit can be in various states. The initial $n_1$ states are classified as minor damage, whereas the subsequent states are categorized as major damage, depending on the extent of the damage.

**Assumption 3**. External shocks unfold in accordance with a PH-renewal process, where the time between two consecutive shocks follows a PH distribution with representation $(\boldsymbol{\gamma}, \mathbf{L})$. The order of $\mathbf{L}$ is equal to $t$.

**Assumption 4**. An external shock has the potential to trigger a complete, non-repairable failure of the online unit, with a probability represented by $\omega^0$.

**Assumption 5**. Following an external shock, the internal performance phase might experience a change. The transition between any two internal states triggered by the external shock is governed by the transition probability matrix $\mathbf{W}$. The column vectors $\mathbf{W}_r^0$ and $\mathbf{W}_{nr}^0$ hold the probabilities of repairable and non-repairable failures, respectively, resulting from an external shock on the internal performance states.

**Assumption 6**. After external shock a cumulative damage is produced. The number of external damage phases is $d$ and the probability transitions between them is given by matrix $\mathbf{D}$. The column vector $\mathbf{D}^0$ contains the probabilities of non-repairable failure from the external cumulative damage phase. The initial distribution for the external cumulative damage when a unit occupies the online place is $\boldsymbol{\omega} = (1,\mathbf{0})$. The cumulative damage phases are partitioned into multiple states, with the first $d_1$ phases considered minor cumulative damage and the remaining phases categorized as major cumulative based on the level of damage.

**Assumption 7**. Random inspections over the online unit can occur while the online place is occupied. The interval between two successive inspections follows a PH distribution with representation $(\boldsymbol{\eta}, \mathbf{M})$ with order $\varepsilon$. This time stops when there are no operational units and restarts when the system transitions from non-operational units to an operational unit.

**Assumption 8**. The duration of vacation time is distributed according to a PH distribution with representation $(\boldsymbol{\upsilon}, \mathbf{V})$, characterized by an order of $v$.

**Assumption 9**. The time required for corrective repair follows a PH distribution $(\boldsymbol{\beta}^1, \mathbf{S}_1)$ with an order equal to $z_1$.

**Assumption 10**. The time needed for preventive maintenance is governed by a PH distribution $(\boldsymbol{\beta}^2, \mathbf{S}_2)$, characterized by an order equal to $z_2$.

Figure 1 shows the behaviour of the system. This figure presents four possible events that may occur. In the first case, the main unit may experience a repairable failure; if this happens, the main unit enters the repair facility, and a standby unit takes its place as the online unit. In any situation, the repairperson may be in the repair facility or on vacation. This repairable failure can occur due to an internal malfunction or an external shock, leading to a change in operation that results in a repairable system failure.

The second quadrant illustrates the occurrence of a non-repairable failure. It is shown that if there are standby units available, the failed unit is removed, and a standby unit takes the primary position. If it is the last unit in the system and a non-repairable failure occurs, the entire system is restarted. A non-repairable failure can occur directly from internal operation or as a result of an external shock. This latter case may occur if the external shock causes total failure, alters the internal behavior leading to non-repairable failure, or exceeds a threshold due to damage from previous external shocks.

The third case illustrates the behavior during inspection. After a phase-type random time, an inspection occurs. If major internal damage is detected (or due to external shocks), the unit is transferred to the repair facility for preventive maintenance. In this case, a standby unit takes the primary position. The repairperson may or may not be present in the repair facility.

Finally, the vacation policy is shown. After a phase-type random period, the repairperson returns from vacation. If the repairperson observes $R$ or more operational units, a new vacation period begins (the system does not appear to be at risk). Otherwise, the repairperson takes their place in the repair facility and begins working. If, upon the repairperson's return, the system has fewer than $R$ units, the repairperson remains in the repair channel. In any case, if there are $R$ or more operational units after a repair, the repairperson begins a new vacation period.

### 2.2.The state-space

The state-space $S$ is composed of macro-states at three concatenated levels.

*First Level*

The first level of the macro-state is $S=\left\{\mathbf{U}^n,\mathbf{U}^{n-1},\ldots,\mathbf{U}^1\right\}$, where $\mathbf{U}^k$ contains the macro-states when there are $k$ units in the system.

*Second level*

Each macro-state $\mathbf{U}^k$ contains the macro-states $\mathbf{E}_s^{k,v}$ and $\mathbf{E}_s^{k,nv}$; $k$ units in the system and $s$ of them in the repair facility,

$$\mathbf{U}^k=\left\{\mathbf{E}_0^{k,v},\mathbf{E}_1^{k,v},...,\mathbf{E}_{N-1}^{k,v},\mathbf{E}_N^{k,v},\mathbf{E}_{N+1}^{k,v},\ldots,\mathbf{E}_k^{k,v},\mathbf{E}_N^{k,nv},\mathbf{E}_{N+1}^{k,nv},\ldots,,\mathbf{E}_k^{k,nv}\right\};\ k\geq R,$$

$$\mathbf{U}^k=\left\{\mathbf{E}_0^{k,nv},\mathbf{E}_1^{k,nv},\ldots,\mathbf{E}_k^{k,nv}\right\};\ k<R.$$

The superscript $v$ and $nv$ indicates whether the repairperson is on vacation or not respectively.

On the one hand, when $k \geq R$ (threshold of the number of operational units for the vacation policy), the repairer can be on vacation for any number of operational units. However, it will be in the repair facility if the number of units in it (corrective repair or preventive maintenance) is greater than or equal to $N = k - R + 1$.

On the other hand, if the number of units' $k$ is less than $R$, the repairperson must always be at his workplace, even interrupting his vacation period if applicable.

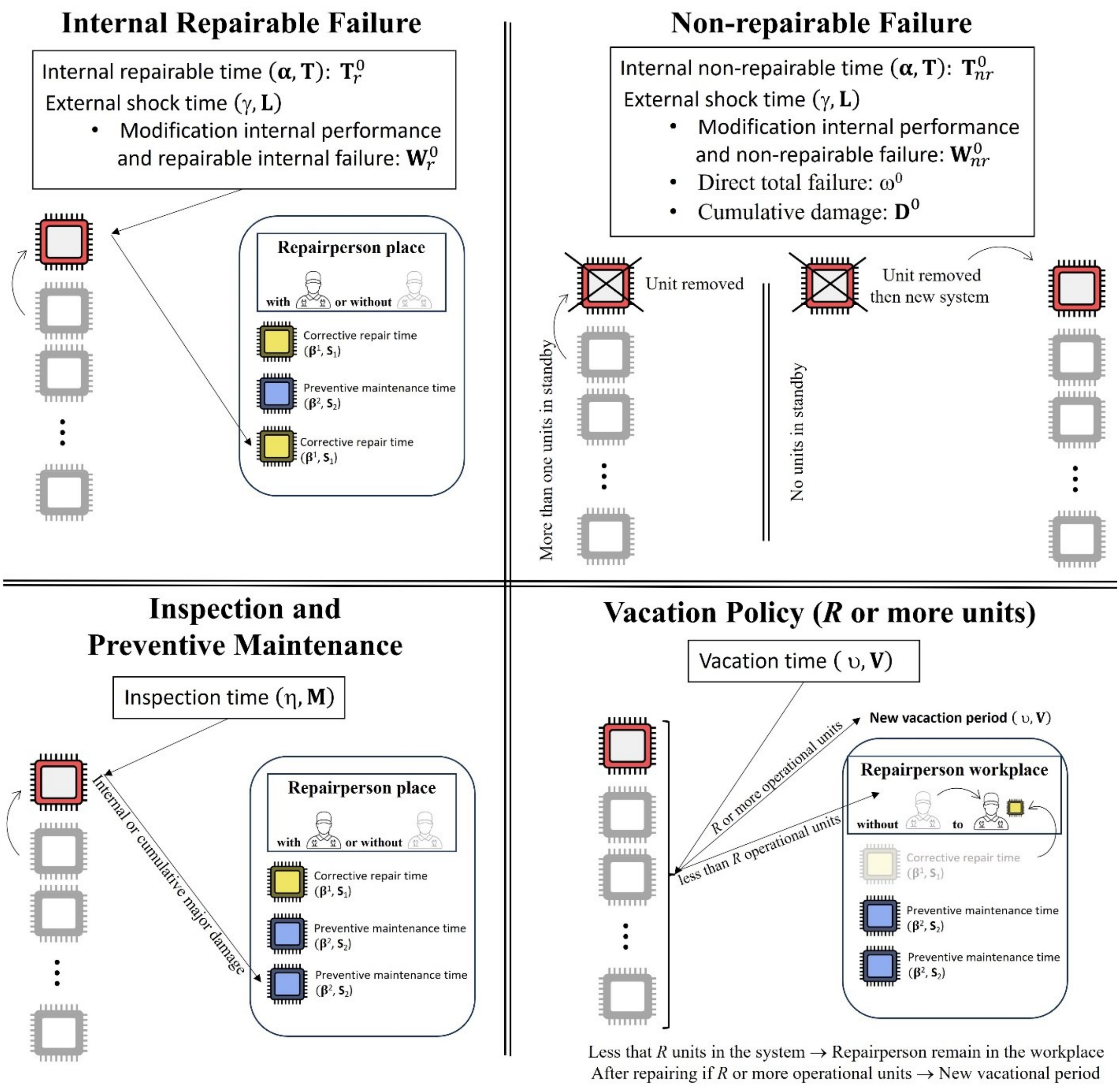

Figure 1. Behaviour of the system: failures, preventive maintenance and vacation polcy

*Third level*

Regardless of the scenario, the sequence of units in the repair facility needs to be retained in memory, with corrective and preventive maintenance being the two repair types. Consequently, the macro-states $\mathbf{E}_s^{k,v}$ and $\mathbf{E}_s^{k,nv}$ are composed of the third level macro-states $\mathbf{E}_{i_1,\dots,i_s}^{k,x}$ being $x = v, nv$.

These macro-states encompass scenarios where there are k units within the system, with $s$ of them currently undergoing repairs at the repair facility. The type of repair is indicated by the ordered sequence $i_1,\dots,i_s$. Specifically, $i_l$ takes on values of 1 or 2 to indicate whether the unit has undergone corrective failure (pending corrective repair) or major inspection (pending preventive maintenance), respectively.

When the quantity of units in the system reaches $R-1$ units, the repairer promptly assumes their work position if he is on vacation.

The phases of this third level macro-state are,

- For $k = 1, \ldots, R-1$

$$\mathbf{E}_0^{k,nv} = \left\{ \left(k,0;i,j,h,u\right); i=1,\ldots,m,\ j=1,\ldots,t, h=1,\ldots,d,\ u=1,\ldots,\varepsilon \right\},$$

$\mathbf{E}_s^{k,nv} = \left\{ \mathbf{E}_{i_1,\ldots,i_s}^{k,nv}; i_l = 1,2; l=1,\ldots,s \right\}$ for $s = 1, \ldots, k$ where

$$\mathbf{E}_{i_1,\ldots,i_s}^{k,nv} = \left\{ \left(k,s;i,j,h,u,r\right); i=1,\ldots,m,\ j=1,\ldots,t, h=1,\ldots,d, u=1,\ldots,\varepsilon, r=1,\ldots,z_{i_1} \right\}$$

for $s < k$ and for $s = k$,

$$\mathbf{E}_{i_1,\ldots,i_k}^{k,nv} = \left\{ \left(k,k;j,r\right); j=1,\ldots,t, u=1,\ldots,\varepsilon, r=1,\ldots,z_{i_1} \right\} .$$

- For $k = R, \ldots, n$

$$\mathbf{E}_0^{k,v} = \left\{ \left(k,0;i,j,h,u,w\right); i=1,\ldots,m,\ j=1,\ldots,t, h=1,\ldots,d, u=1,\ldots,\varepsilon, w=1,\ldots,v \right\},$$

$\mathbf{E}_s^{k,v} = \left\{ \mathbf{E}_{i_1,\ldots,i_s}^{k,v}; i_l = 1,2; l=1,\ldots,s \right\}$ for $s = 1, \ldots, k$ where

$$\mathbf{E}_{i_1,\ldots,i_s}^{k,v} = \left\{ \left(k,s;i,j,h,u,w\right); i=1,\ldots,m,\ j=1,\ldots,t, h=1,\ldots,d, u=1,\ldots,\varepsilon, w=1,\ldots,v \right\}$$

for $s < k$ and for $s = k$,

$$\mathbf{E}_{i_1,\ldots,i_k}^{k,v} = \left\{ \left(k,k;j,w\right); j=1,\ldots,t, w=1,\ldots,v \right\}.$$

$\mathbf{E}_s^{k,nv} = \left\{ \mathbf{E}_{i_1,\ldots,i_s}^{k,nv}; i_l = 1,2; l=1,\ldots,s \right\}$ for $s = N, \ldots, k$ where

$$\mathbf{E}_{i_1,\ldots,i_s}^{k,nv} = \left\{ \left(k,s;i,j,h,u,r\right); i=1,\ldots,m,\ j=1,\ldots,t, h=1,\ldots,d, u=1,\ldots,\varepsilon, r=1,\ldots,z_{i_1} \right\}$$

for $s < k$ and for $s = k$,

$$\mathbf{E}_{i_1,\ldots,i_k}^{k,nv} = \left\{ \left(k,k;j,r\right); j=1,\ldots,t, r=1,\ldots,z_{i_1} \right\} .$$

The phase ($k, s; i, j, h, u, r$) signifies various aspects of the system's state:

- $k$ units are within the system, with $s$ units currently in the maintenance center.
- The online unit's internal performance is represented by state $i$.
- The state of the external shock time is denoted by $j$.
- $h$ denotes the cumulative external damage.
- $u$ is the inspection time phase.

- $r$ indicates the corrective repair or preventive maintenance phase for the unit currently undergoing service at the repair facility.
- $w$ indicates the phase of the repairperson's vacation, if applicable.

## 3. The model. Markovian Arrival Process with Marked Arrivals

The system is regulated by a continuous-time Markov process vector. Within this section, in order to build a model for the envisioned complex system, the model of the online unit and after the MMAP is built in detail.

### 3.1. Modelling the Online Unit

The online unit may experience various types of events at any given time. These events are categorized as follows:

$A$: Repairable failure
$B$: Positive inspection (major inspection)
$C$: Non-repairable failure from internal performance or after external shock
$O$: No events

The attention is focused on the transition operating to non-repairable failure, that is matrix $\mathbf{H}_C$ and, on the transition, operating to major damage, matrix $\mathbf{H}_B$. Through this work, $\mathbf{e}$ is a vector of ones with appropriate order and, the symbol $\otimes$ denotes the well-known Kronecker product. The rest of matrices are given in Appendix A.

Matrix $\mathbf{H}_C$

This matrix block contains the transition intensities from the operational phases of the online unit to a non-repairable failure.

The non-repairable failure can be provoked by the following events:

- Internal non repairable failure. It occurs through the column vector $\mathbf{T}_{nr}^{0}$, following a new unit occupies the online place with initial distribution $\boldsymbol{\alpha}$ for the internal performance. Given that only one transition can occur at any time there is not an external shock, the new online unit begins the external damage with matrix $\mathbf{e}\boldsymbol{\omega}$ and a new inspection period time begins $\mathbf{e}\boldsymbol{\eta}$. Then,

$$\mathbf{T}_{nr}^{0}\boldsymbol{\alpha}\otimes\mathbf{I}\otimes\mathbf{e}\boldsymbol{\omega}\otimes\mathbf{e}\boldsymbol{\eta}.$$

- External shock and non-repairable failure from modification of the internal behaviour. An external shock is produced without extreme failure and a new external shock time begins, $\mathbf{L}^{0}\boldsymbol{\gamma}\left(1-\omega^{0}\right)$. This external shock modifies the internal structure of the online unit by producing a non-repairable failure and the new online unit is reinitialized, $\mathbf{W}_{nr}^{0}\boldsymbol{\alpha}$. Finally, the external shock does not provoke a non-repairable failure due to cumulative external damage and this cumulative external damage is restarted for the new online unit, $\mathbf{D}\mathbf{e}\boldsymbol{\omega}$. A new inspection period time begins $\mathbf{e}\boldsymbol{\eta}$. Then,

$$\mathbf{W}_{nr}^{0}\boldsymbol{\alpha}\otimes\mathbf{L}^{0}\boldsymbol{\gamma}\left(1-\omega^{0}\right)\otimes\mathbf{De\omega}\otimes\mathbf{e\eta}.$$

- Non-repairable failure due to cumulative external damage. An external shock is produced without extreme failure and a new external shock time begins, $\mathbf{L}^{0}\boldsymbol{\gamma}\left(1-\omega^{0}\right)$. This external shock provokes a non-repairable failure due to cumulative external damage and this cumulative external damage is restarted for the new online unit, $\mathbf{De\omega}$. A new inspection period time begins $\mathbf{e\eta}$. Then,

$$\mathbf{e\alpha}\otimes\mathbf{L}^{0}\boldsymbol{\gamma}\left(1-\omega^{0}\right)\otimes\mathbf{D}^{0}\boldsymbol{\omega}\otimes\mathbf{e\eta}.$$

- Non-repairable failure due to total extreme failure after an external shock. The cumulative external damage is restarted for the new online unit, $\mathbf{e\omega}$, and a new inspection time begins, $\mathbf{e\eta}$

$$\mathbf{e\alpha}\otimes\mathbf{L}^{0}\boldsymbol{\gamma}\omega^{0}\otimes\mathbf{e\omega}\otimes\mathbf{e\eta}.$$

If there are at least one operational unit in standby, after a non-repairable failure this one occupies the online place and a new random inspection time begins. Therefore the matrix $\mathbf{H}_C$ is

$$\mathbf{H}_{C}=\left[\mathbf{T}_{nr}^{0}\boldsymbol{\alpha}\otimes\mathbf{I}\otimes\mathbf{e\omega}+\mathbf{W}_{nr}^{0}\boldsymbol{\alpha}\otimes\mathbf{L}^{0}\boldsymbol{\gamma}\left(1-\omega^{0}\right)\otimes\mathbf{De\omega}\right.$$
$$\left.+\mathbf{e\alpha}\otimes\mathbf{L}^{0}\boldsymbol{\gamma}\left(1-\omega^{0}\right)\otimes\mathbf{D}^{0}\boldsymbol{\omega}+\mathbf{e\alpha}\otimes\mathbf{L}^{0}\boldsymbol{\gamma}\omega^{0}\otimes\mathbf{e\omega}\right]\otimes\mathbf{e\eta}.$$

If there is only one operational unit, then this matrix is replaced by

$$\mathbf{H}_{C}^{'}=\left[\mathbf{T}_{nr}^{0}\otimes\mathbf{I}\otimes\mathbf{e}+\mathbf{W}_{nr}^{0}\otimes\mathbf{L}^{0}\boldsymbol{\gamma}\left(1-\omega^{0}\right)\otimes\mathbf{De}\right.$$
$$\left.+\mathbf{e}\otimes\mathbf{L}^{0}\boldsymbol{\gamma}\left(1-\omega^{0}\right)\otimes\mathbf{D}^{0}+\mathbf{e}\otimes\mathbf{L}^{0}\boldsymbol{\gamma}\omega^{0}\otimes\mathbf{e}\right]\otimes\mathbf{e}.$$

Matrix $\mathbf{H}_B$

This matrix block contains the transition intensities from the operational phases of the online unit to a major damage event observed during inspection of the online unit.

To model the behavior of minor and major damage after inspection for internal performance and external damage, the matrices $\mathbf{U}_i$ and $\mathbf{V}_i$ are defined respectively as follows:

Minor internal damage: $U_1\left(i,j\right)=\begin{cases}1 & ;\quad i=j; i=1,\dots,n_1\\ 0 & ;\quad \text{otherwise}\end{cases}$,

Mayor internal damage: $U_2\left(i,j\right)=\begin{cases}1 & ;\quad i=j; i=n_1,\dots,m\\ 0 & ;\quad \text{otherwise}\end{cases}$,

Minor cumulative damage due to external shock: $V_1(i,j)=\begin{cases}1 & ; \quad i=j; i=1,...,d_1 \\ 0 & ; \quad \text{otherwise}\end{cases}$,

Mayor cumulative damage due to external shock: $V_2(i,j)=\begin{cases}1 & ; \quad i=j; i=d_1,...,d \\ 0 & ; \quad \text{otherwise}\end{cases}$.

When these matrices pre-multiply another matrix, the only non-zero transitions that result are those indicating the event occurrence.

Inspection observes major damage in the main unit when any of the following situations occur:

- An inspection takes place ($\mathbf{M}^0\boldsymbol{\eta}$), and major internal damage is observed in the main unit ($\mathbf{U}_2\mathbf{e}\boldsymbol{\alpha}$) causing the unit to enter the repair facility and a reserve unit to become operational. Since major internal damage has been observed and the unit enters preventive maintenance, it is considered that major or minor damage due to external shock may also have been observed, $(\mathbf{V}_1+\mathbf{V}_2)\mathbf{e}\boldsymbol{\omega}=\mathbf{e}\boldsymbol{\omega}$. This unit starts with the initial distribution for external shock damage. Given that infinitesimally only one unconditioned event can occur, the transition intensity for external shock damage does not change (**I**).
$$\mathbf{U}_2\mathbf{e}\boldsymbol{\alpha}\otimes\mathbf{I}\otimes\mathbf{e}\boldsymbol{\omega}\otimes\mathbf{M}^0\boldsymbol{\eta}.$$
- The second possible situation is as follows. An inspection takes place ($\mathbf{M}^0\boldsymbol{\eta}$), and no major internal damage is observed in the main unit ($\mathbf{U}_1\mathbf{e}\boldsymbol{\alpha}$), but major damage caused by external shock is detected ($\mathbf{V}_2\mathbf{e}\boldsymbol{\omega}$), causing the unit to enter the repair facility and a reserve unit to become operational with the corresponding initial distribution for external shock damage. Given that infinitesimally only one unconditioned event can occur, the transition intensity for external shock damage does not change (**I**). That is,
$$\mathbf{U}_1\mathbf{e}\boldsymbol{\alpha}\otimes\mathbf{I}\otimes\mathbf{V}_2\mathbf{e}\boldsymbol{\omega}\otimes\mathbf{M}^0\boldsymbol{\eta}.$$

Therefore, we have $\mathbf{H}_B=\mathbf{U}_2\mathbf{e}\boldsymbol{\alpha}\otimes\mathbf{I}\otimes\mathbf{e}\boldsymbol{\omega}\otimes\mathbf{M}^0\boldsymbol{\eta}+\mathbf{U}_1\mathbf{e}\boldsymbol{\alpha}\otimes\mathbf{I}\otimes\mathbf{V}_2\mathbf{e}\boldsymbol{\omega}\otimes\mathbf{M}^0\boldsymbol{\eta}$,

It could happen that when the inspected main unit enters preventive maintenance, there are no reserve units available. In this case, the initial distribution for internal operation ($\boldsymbol{\alpha}$) and external wear damage ($\boldsymbol{\omega}$) should not be considered. In this scenario, we would have

$$\mathbf{H}_B^{'}=\mathbf{U}_2\mathbf{e}\otimes\mathbf{I}\otimes\mathbf{e}\otimes\mathbf{M}^0+\mathbf{U}_1\mathbf{e}\otimes\mathbf{I}\otimes\mathbf{V}_2\mathbf{e}\otimes\mathbf{M}^0.$$

In Appendix A are given the rest of the matrices for the online unit.

### 3.2. The Markovian Arrival Process with marked arrivals (MMAP)

The system's behavior is regulated by a **MMAP** (Marked Markovian Arrival Process). The representation of this **MMAP** is derived from the types of events described below:

*A*: Repairable failure

*B*: Positive inspection

*C*: Non-repairable failure (without returning to work)

*D*: Returns to work after vacation (without non-repairable failure)

*CD*: Non-repairable failure and return to work (interrupt the vacation)

*E*: The repairperson arrives and new vacation period

*F*: New vacation period after repair

*NS*: New system and therefore a new vacation period begins.

*O*: No events

The representation of the **MMAP** is $\left(\mathbf{D}^O, \mathbf{D}^A, \mathbf{D}^B, \mathbf{D}^C, \mathbf{D}^D, \mathbf{D}^{CD}, \mathbf{D}^E, \mathbf{D}^F, \mathbf{D}^{NS}\right)$.

The *generator* corresponding to the embedded Markov chain within the **MMAP** is provided as follows: $\mathbf{D} = \sum_Y \mathbf{D}^Y$ .

The matrix $\mathbf{D}^C$ is described below and the rest are given in Appendix B.

*The matrix* $\mathbf{D}^C$

The matrix $\mathbf{D}^C$ is composed of the transition intensities when only a non-repairable failure occurs. It is an upper diagonal matrix block because the number of units in the system changes during these transitions. This matrix is composed of matrix blocks corresponding to changes between macro-states defined at first level, $\mathbf{U}^k$.

$$\mathbf{D}^C = \begin{pmatrix} \mathbf{0} & \mathbf{D}_n^C & & & \\ & \mathbf{0} & \mathbf{D}_{n-1}^C & & \\ & & \mathbf{0} & \ddots & \\ & & & \ddots & \mathbf{D}_2^C \\ \mathbf{0} & & & & \mathbf{0} \end{pmatrix}.$$

The matrix block $\mathbf{D}_k^C$ contains matrix blocks for the transitions between macro-states of the second level, $E_s^{k,x}$ (*k* units within the system, with *s* of them undergoing repairs at the repair facility).

If the number of units in the system is less than *R*, the repairperson is always in the repair facility. Then, for $k = 2, \ldots, R-1$ and $k \neq R \geq 3$

$$
\mathbf{D}_k^C = \begin{array}{c} \\ E_0^{k,nv} \\ E_1^{k,nv} \\ \vdots \\ E_{k-1}^{k,vn} \\ E_k^{k,nv} \end{array}
\begin{array}{c} \begin{array}{cccc} E_0^{k-1,nv} & E_1^{k-1,nv} & \cdots & E_{k-1}^{k-1,nv} \end{array} \\
\left( \begin{array}{cccc} \mathbf{D}_{00}^{C,k,nv} & & & \\ & \mathbf{D}_{11}^{C,k,nv} & & \\ & & \ddots & \\ & & & \mathbf{D}_{k-1,k-1}^{C,k,nv} \\ & & & \mathbf{0} \end{array} \right) \end{array}.
$$

If the number of units is exactly $R$, the repairperson could be on vacation before the transition but after a non-repairable failure the repairperson remains in the repair facility.

For $k = R \geq 2$,

$$
\mathbf{D}_k^C = \begin{array}{c} \\ E_0^{k,v} \\ E_1^{k,v} \\ \vdots \\ E_{k-1}^{k,v} \\ E_k^{k,v} \\ E_{N=1}^{k,nv} \\ E_2^{k,nv} \\ \vdots \\ E_{k-1}^{k,nv} \\ E_k^{k,nv} \end{array}
\begin{array}{c} \begin{array}{cccc} E_0^{k-1,nv} & E_1^{k-1,nv} \cdots & E_{k-2}^{k-1,nv} & E_{k-1}^{k-1,nv} \end{array} \\
\left( \begin{array}{cccc} \mathbf{0} & & & \mathbf{0} \\ & & & \\ & \ddots & & \\ & & & \\ \mathbf{0} & & & \mathbf{0} \\ \hdashline \mathbf{0} & \mathbf{D}_{11}^{C,k,nv} & & \\ & \mathbf{0} & \ddots & \\ & & \ddots & \ddots \\ & & \mathbf{0} & \mathbf{D}_{k-1,k-1}^{C,k,nv} \\ \mathbf{0} & \cdots & \cdots & \mathbf{0} \end{array} \right) \end{array}
$$

Finally, if the number of units in the system exceeds $R$, the repairer may either be at the repair facility or on vacation before and after of the transition.

For $k = R+1,\ldots, n$ with $R \leq n-1$

$$
\mathbf{D}_k^C = \begin{array}{c} \\ E_0^{k,v} \\ E_1^{k,v} \\ \vdots \\ E_{N-1}^{k,v} \\ E_N^{k,v} \\ E_{N+1}^{k,v} \\ \vdots \\ E_{k-1}^{k,v} \\ E_k^{k,v} \\ E_N^{k,nv} \\ E_{N+1}^{k,nv} \\ \vdots \\ E_{k-1}^{k,nv} \\ E_k^{k,nv} \end{array}
\begin{array}{ccccccccc|ccccc}
E_0^{k-1,v} & E_1^{k-1,v} & \cdots & E_{N-1}^{k-1,v} & E_N^{k-1,v} & E_{N+1}^{k-1,v} & \cdots & E_{k-2}^{k-1,v} & E_{k-1}^{k-1,v} & E_{N-1}^{k-1,nv} & E_N^{k-1,nv} & \cdots & E_{k-2}^{k-1,nv} & E_{k-1}^{k-1,nv} \\
\mathbf{D}_{00}^{C,k,v} & & & & & & & & & & & & & \\
& \mathbf{D}_{11}^{C,k,v} & & & & & & & & & & & & \\
& & \ddots & & & & & & & & & & & \\
& & & \mathbf{D}_{N-1,N-1}^{C,k,v} & & & & & & & & & & \\
& & & & \mathbf{D}_{N,N}^{C,k,v} & & & & & & & & & \\
& & & & & \mathbf{D}_{N+1,N+1}^{C,k,v} & & & & & & & & \\
& & & & & & \ddots & & & & & & & \\
& & & & & & & & \mathbf{D}_{k-1,k-1}^{C,k,v} & & & & & \\
& & & & & & & & \mathbf{0} & & & & & \\
\hline
& & & & & & & & & \mathbf{0} & \mathbf{D}_{N,N}^{C,k,nv} & & & \\
& & & & & & & & & & \mathbf{0} & \mathbf{D}_{N+1,N+1}^{C,k,nv} & & \\
& & & & & & & & & & & \ddots & \ddots & \\
& & & & & & & & & & & & \mathbf{0} & \mathbf{D}_{k-1,k-1}^{C,k,nv} \\
& & & & & & & & & & & & & \mathbf{0}
\end{array}
$$

In all cases, the block $\mathbf{D}_{rr}^{C,k,x}$ contains the transition intensities from $k$ units in the system to $k$–1 units in the system when there are $r$ units in the repair facility, where $x$ is equal to $v$ or $nv$ depending on the repairperson is on vacation or not respectively.

These matrices are built by considering the double possibility of task for the repair technician, encompassing both corrective repairs and preventive maintenance.

When all units are operational, $\mathbf{D}_{00}^{C,k,nv} = \mathbf{H}_C$.

Throughout the paper, the indicator function $I_{\{\cdot\}}$ is used, taking the value one if the condition within the brackets is true, or zero otherwise.

For $r = 0, \ldots, k-1$, $\mathbf{D}_{r,r}^{C,k,v} = \mathbf{I}_{2^r} \otimes \left( I_{\{r=k-1\}} \mathbf{H}'_C + I_{\{r<k-1\}} \mathbf{H}_C \right) \otimes \mathbf{I}$.

For $r$ =1, …, k–1, $\mathbf{D}_{r,r}^{C,k,nv} = \begin{pmatrix} \mathbf{CR}_{r,r}^{C,k,nv} \\ \mathbf{PM}_{r,r}^{C,k,nv} \end{pmatrix}$, where the matrix

$\mathbf{CR}_{r,r}^{C,k,nv} = \left( \mathbf{I}_{2^{r-1}} \otimes \left( I_{\{r=k-1\}} \mathbf{H}'_C + I_{\{r<k-1\}} \mathbf{H}_C \right) \otimes \mathbf{I} \quad \mathbf{0} \right)$ contains the transition to non-repairable failure when the repairperson is working on corrective repair, and $\mathbf{PM}_{r,r}^{C,k,nv} = \left( \mathbf{0} \quad \mathbf{I}_{2^{r-1}} \otimes \left( I_{\{r=k-1\}} \mathbf{H}'_C + I_{\{r<k-1\}} \mathbf{H}_C \right) \otimes \mathbf{I} \right)$ when the repairperson is working on preventive maintenance. In both cases, each column indicates corrective repair and preventive maintenance respectively.

## 4. Measures

This section provides descriptions of various interesting measures that can be calculated in both the transient and stationary regimen.

### 4.1. The transient and the stationary distribution

The transient distribution is calculated by using the initial distribution and the generator matrix of the vector Markov process defined from the MMAP outlined in Section 3.2.

The initial distribution of the Markov process is represented as $\boldsymbol{\phi} = \left(\boldsymbol{\alpha} \otimes \boldsymbol{\gamma}_{st} \otimes \boldsymbol{\eta} \otimes \boldsymbol{\omega}, \mathbf{0}\right)$, where $\boldsymbol{\gamma}_{st}$ is the stationary distribution of the phase-type renewal process with a matrix generator denoted as $\mathbf{L} + \mathbf{L}^0\boldsymbol{\gamma}$. As is well known, the stationary distribution satisfies that $\boldsymbol{\gamma}_{st}\left(\mathbf{L} + \mathbf{L}^0\boldsymbol{\gamma}\right) = \mathbf{0}$ and $\boldsymbol{\gamma}_{st}\mathbf{e} = 1$. If $\mathbf{M}^*$ denotes a matrix $\mathbf{M}$ with the first column removed, the aforementioned conditions can be expressed jointly as $\boldsymbol{\gamma}_{st}\left[\mathbf{e} \,|\, \left(\mathbf{L} + \mathbf{L}^0\boldsymbol{\gamma}\right)^*\right] = (1, \mathbf{0})$. Consequently, $\boldsymbol{\gamma}_{st} = (1, \mathbf{0})\left(\mathbf{e} \,|\, \left(\mathbf{L} + \mathbf{L}^0\boldsymbol{\gamma}\right)^*\right)^{-1}$.

The probability of being in the macro-states $\mathbf{E}_s^{k,x}$ at time $t$ is calculated by using matrix blocks as $\mathbf{p}_{\mathbf{E}_s^{k,x}}(t) = \left(\boldsymbol{\phi} e^{\mathbf{D}t}\right)_{I_s^{k,x}}$ where $I_s^{k,x}$ denotes the range for the corresponding states. Clearly, vector $\mathbf{p}(t) = \boldsymbol{\phi} e^{\mathbf{D}t}$ represents the transitory distribution at time *t*.

*The stationary distribution (steady-state)*

To compute the stationary distribution in a matrix-algorithmic way, the matrix **D** is divided into the following distinct blocks (according to the macro-states $\mathbf{U}^j$),

$$\mathbf{D} = \begin{pmatrix} \mathbf{D}_{n,n} & \mathbf{D}_{n,n-1} & \mathbf{0} & \cdots & \mathbf{0} & \mathbf{0} \\ \mathbf{0} & \mathbf{D}_{n-1,n-1} & \mathbf{D}_{n-1,n-2} & \cdots & \mathbf{0} & \mathbf{0} \\ \vdots & \vdots & \vdots & \ddots & \vdots & \vdots \\ \mathbf{0} & \mathbf{0} & \cdots & \cdots & \mathbf{D}_{22} & \mathbf{D}_{21} \\ \mathbf{D}_{1n} & \mathbf{0} & \cdots & \cdots & & \mathbf{D}_{11} \end{pmatrix},$$

where for $i = 1, \ldots, n$

$$\mathbf{D}_{ii} = \mathbf{D}_i^O + \mathbf{D}_i^A + \mathbf{D}_i^B + \mathbf{D}_i^D + \mathbf{D}_i^E + \mathbf{D}_i^F,$$

and for $i = 2, \ldots, n$

$$\mathbf{D}_{i,i-1} = \mathbf{D}_i^C + I_{\{i=R\geq 2\}}\mathbf{D}_i^{CD} \text{ and } \mathbf{D}_{1n} = \mathbf{D}_1^{NS}.$$

The stationary distribution $\boldsymbol{\pi}$ can be partitioned into blocks as $\boldsymbol{\pi}=\left(\boldsymbol{\pi}_n,\boldsymbol{\pi}_{n-1},\ldots,\boldsymbol{\pi}_1\right)$ and it verifies the balance equations which can be expressed in a matrix form as $\boldsymbol{\pi}\mathbf{D}=\mathbf{0}$ and $\boldsymbol{\pi}\mathbf{e}=1$. That is,

$$\boldsymbol{\pi}_n\mathbf{D}_{n,n}+\boldsymbol{\pi}_1\mathbf{D}_{1n}=\mathbf{0},$$

$$\boldsymbol{\pi}_n\mathbf{D}_{n,n-1}+\boldsymbol{\pi}_{n-1}\mathbf{D}_{n-1,n-1}=\mathbf{0},$$

$$\boldsymbol{\pi}_{n-1}\mathbf{D}_{n-1,n-2}+\boldsymbol{\pi}_{n-2}\mathbf{D}_{n-2,n-2}=\mathbf{0},$$

$$\ldots$$

$$\boldsymbol{\pi}_3\mathbf{D}_{32}+\boldsymbol{\pi}_2\mathbf{D}_{22}=\mathbf{0},$$

$$\boldsymbol{\pi}_2\mathbf{D}_{21}+\boldsymbol{\pi}_1\mathbf{D}_{11}=\mathbf{0},$$

$$\boldsymbol{\pi}\mathbf{e}=1.$$

This matrix system has been solved by matrix blocks and it can be expressed as

$$\boldsymbol{\pi}_n=-\boldsymbol{\pi}_1\mathbf{D}_{1n}\mathbf{D}_{n,n}^{-1}$$

and for $k = 2,\ldots, n-1$

$$\boldsymbol{\pi}_k=\left(-1\right)^{I_{\{n=\dot{2}\}}(k+1)+I_{\{n\neq\dot{2}\}}k}\,\boldsymbol{\pi}_1\mathbf{D}_{1n}\mathbf{D}_{n,n}^{-1}\prod_{i=0}^{n-k-1}\mathbf{G}_{n-i,n-i-1},$$

being

$$\mathbf{G}_{i,i-1}=\mathbf{D}_{i,i-1}\mathbf{D}_{i-1,i-1}^{-1}\ \text{for } i = 3,\ldots,n$$

and finally $\boldsymbol{\pi}\mathbf{e}=1$, then

$$\sum_{k=1}^{n}\boldsymbol{\pi}_k\mathbf{e}=\boldsymbol{\pi}_1\mathbf{e}+\boldsymbol{\pi}_1\sum_{k=2}^{n-1}\left(-1\right)^{I_{\{n=\dot{2}\}}(k+1)+I_{\{n\neq\dot{2}\}}k}\,\mathbf{D}_{1n}\mathbf{D}_{n,n}^{-1}\prod_{i=0}^{n-k-1}\mathbf{G}_{n-i,n-i-1}\mathbf{e}-\boldsymbol{\pi}_1\mathbf{D}_{1n}\mathbf{D}_{n,n}^{-1}\mathbf{e}=1,$$

$$\begin{aligned}\boldsymbol{\pi}_1\Bigg[&\left(\left(-1\right)^{I_{\{n=\dot{2}\}}+2I_{\{n\neq\dot{2}\}}}\,\mathbf{D}_{1n}\mathbf{D}_{n,n}^{-1}\prod_{i=0}^{n-3}\mathbf{G}_{n-i,n-i-1}\mathbf{D}_{21}+\mathbf{D}_{11}\right)^{*}\Bigg|\\ &\mathbf{e}+\mathbf{D}_{1n}\mathbf{D}_{n,n}^{-1}\sum_{k=2}^{n-1}\left(-1\right)^{I_{\{n=\dot{2}\}}(k+1)+I_{\{n\neq\dot{2}\}}k}\prod_{i=0}^{n-k-1}\mathbf{G}_{n-i,n-i-1}\mathbf{e}-\mathbf{D}_{1n}\mathbf{D}_{n,n}^{-1}\mathbf{e}\Bigg]=\left(\mathbf{0},1\right),\end{aligned}$$

and therefore from this expression and $\boldsymbol{\pi}_2\mathbf{D}_{21}+\boldsymbol{\pi}_1\mathbf{D}_{11}=\mathbf{0}$,

$$\boldsymbol{\pi}_1 = (\mathbf{0},1)\left[\left((-1)^{I_{\{n=2\}}+2I_{\{n\neq 2\}}}\mathbf{D}_{1n}\mathbf{D}_{n,n}^{-1}\prod_{i=0}^{n-3}\mathbf{G}_{n-i,n-i-1}\mathbf{D}_{21}+\mathbf{D}_{11}\right)^{*}\right|$$

$$\left.\mathbf{e}+\mathbf{D}_{1n}\mathbf{D}_{n,n}^{-1}\sum_{k=2}^{n-1}(-1)^{I_{\{n=2\}}(k+1)+I_{\{n\neq 2\}}k}\prod_{i=0}^{n-k-1}\mathbf{G}_{n-i,n-i-1}\mathbf{e}-\mathbf{D}_{1n}\mathbf{D}_{n,n}^{-1}\mathbf{e}\right]^{-1}.$$

### 4.2. Availability, mean times in macro-states and mean operational time

The availability is the probability of being operational the system at a certain time $t$. The system is operational when at least one unit is operational, that is

$$A(t)=1-\sum_{k=R}^{n}\mathbf{p}_{E_k^{k,v}}(t)\mathbf{e}-\sum_{k=1}^{n}\mathbf{p}_{E_k^{k,nv}}(t)\mathbf{e}. \tag{1}$$

The availability in stationary regime is given from (1) by

$$A=1-\sum_{k=R}^{n}\boldsymbol{\pi}_{E_k^{k,v}}\mathbf{e}-\sum_{k=1}^{n}\boldsymbol{\pi}_{E_k^{k,nv}}\mathbf{e}. \tag{2}$$

It is of interest to work out the mean time that the system is at a certain macro-state. The $k$ units in the system case, with $s$ in the repair facility up to a certain time $t$ is equal to,

$$\Psi_{k,s}^{v}(t)=\int_0^t\mathbf{p}_{E_s^{k,v}}(u)du\cdot\mathbf{e}\quad;\quad\Psi_{k,s}^{nv}(t)=\int_0^t\mathbf{p}_{E_s^{k,nv}}(u)du\cdot\mathbf{e}, \tag{3}$$

for the case when the repairperson is either on vacation or in the repair facility respectively, where $s \le k$.

The proportional time that the system is in macro-state $E_s^{k,v}$ and $E_s^{k,nv}$ up to a certain time $t$ can be calculated from (3), and they are given by $\Psi_{k,s}^{v}(t)/t$ and $\Psi_{k,s}^{nv}(t)/t$, respectively. In stationary regime these measures would be

$$\lim_{t\to\infty}\frac{\Psi_{k,s}^{v}(t)}{t}=\Psi_{k,s}^{v}=\boldsymbol{\pi}_{E_s^{k,v}}\mathbf{e}\quad;\quad\lim_{t\to\infty}\frac{\Psi_{k,s}^{nv}(t)}{t}=\Psi_{k,s}^{nv}=\boldsymbol{\pi}_{E_s^{k,nv}}\mathbf{e}. \tag{4}$$

From these measures given in (3), the mean time with $k$ units and $s$ in the repair facility up to time $t$ can be expressed as

$$\Psi_{k,s}(t)=I_{\{(1\le k\le R-1\ \&\ 0\le s\le k)\text{ or }(R\le k\le n\ \&\ k-R+1\le s\le k)\}}\Psi_{k,s}^{nv}(t)+I_{\{R\le k\le n\ \&\ 0\le s\le k\}}\Psi_{k,s}^{v}(t). \tag{5}$$

It is immediate to obtain the proportional time in stationary regime from (4) and (5),

$$\Psi_{k,s}=I_{\{(1\le k\le R-1\ \&\ 0\le s\le k)\text{ or }(R\le k\le n\ \&\ k-R+1\le s\le k)\}}\Psi_{k,s}^{nv}+I_{\{R\le k\le n\ \&\ 0\le s\le k\}}\Psi_{k,s}^{v}.$$

Finally, this measure is determined for the first level of macro-states. The mean time with $k$ units in the system up to a certain time $t$ is

$$\Psi_k(t) = I_{\{1\le k\le R-1\}}\sum_{s=0}^{k}\Psi_{k,s}^{nv}(t) + I_{\{k\ge R\}}\sum_{s=0}^{k}\Psi_{k,s}^{v}(t) + I_{\{k\ge R\}}\sum_{s=k-R+1}^{k}\Psi_{k,s}^{nv}(t).$$

In stationary regime, the proportional time can be calculated from the expression above or directly from the stationary distribution developed in section 4.1. Then,

$$\Psi_k = I_{\{1\le k\le R-1\}}\sum_{s=0}^{k}\Psi_{k,s}^{nv} + I_{\{k\ge R\}}\sum_{s=0}^{k}\Psi_{k,s}^{v} + I_{\{k\ge R\}}\sum_{s=k-R+1}^{k}\Psi_{k,s}^{nv} = \boldsymbol{\pi}_k\mathbf{e}.$$

Thanks to the functions defined in (5), the mean operational time up to a certain time $t$ can be calculated as $\mu_{op}(t) = \sum_{k=1}^{n}\sum_{s=0}^{k-1}\Psi_{k,s}(t)$.

## 4.3. Mean number of events

The system has been modeled by considering a structure that allows for the calculation of the mean number of events defined by the MMAP (proportional number of events per unit of time in stationary regime). Table 1 shows this measure for both, the transient and stationary cases.

| Mean number | Transient (up to time *t*) | Stationary regime |
|---|---|---|
| Mean number of repairable failures | $\Lambda^{rep}(t) = \int_0^t \mathbf{p}(u)du \cdot \mathbf{D}^A \cdot \mathbf{e}$ | $\Lambda^{rep} = \boldsymbol{\pi}\mathbf{D}^A\mathbf{e}$ |
| Mean number of major inspections | $\Lambda^{mi}(t) = \int_0^t \mathbf{p}(u)du \cdot \mathbf{D}^B \cdot \mathbf{e}$ | $\Lambda^{mi} = \boldsymbol{\pi}\mathbf{D}^B\mathbf{e}$ |
| Mean number of non-repairable failures | $\Lambda^{nr}(t) = \int_0^t \mathbf{p}(u)du\left(\mathbf{D}^C + \mathbf{D}^{CD} + \mathbf{D}^{NS}\right)\mathbf{e}$ | $\Lambda^{nr} = \boldsymbol{\pi}\left(\mathbf{D}^C + \mathbf{D}^{CD} + \mathbf{D}^{NS}\right)\mathbf{e}$ |
| Mean number of returns to work | $\Lambda^{ret}(t) = \int_0^t \mathbf{p}(u)du \cdot \left(\mathbf{D}^D + \mathbf{D}^{CD}\right)\cdot \mathbf{e}$ | $\Lambda^{ret} = \boldsymbol{\pi}\left(\mathbf{D}^D + \mathbf{D}^{CD}\right)\mathbf{e}$ |
| Mean number of returns to work and start of a new vacation period | $\Lambda^{ret-be}(t) = \int_0^t \mathbf{p}(u)du \cdot \mathbf{D}^E \cdot \mathbf{e}$ | $\Lambda^{ret-be} = \boldsymbol{\pi}\mathbf{D}^E\mathbf{e}$ |
| Mean number of vacation periods after repair | $\Lambda^{after}(t) = \int_0^t \mathbf{p}(u)du \cdot \mathbf{D}^F \cdot \mathbf{e}$ | $\Lambda^{after} = \boldsymbol{\pi}\mathbf{D}^F\mathbf{e}$ |
| Mean number of new systems | $\Lambda^{NS}(t) = \int_0^t \mathbf{p}(u)du \cdot \mathbf{D}^{NS} \cdot \mathbf{e}$ | $\Lambda^{NS} = \boldsymbol{\pi}\mathbf{D}^{NS}\mathbf{e}$ |

Table 1. Mean number of events up to a certain time and in stationary regime

## 4.4. Rewards and costs

To assess the economic viability of the system, an analysis involving costs and rewards has been conducted. A net profit vector, linked to the state-space, has been constructed and measures worked out.

### 4.4.1. Building the cost/reward vector

Several values associated to rewards and costs while system is or not operational have been introduced:

$B$: gross profit while the system is operational per unit of time.

$\mathbf{c}_0$: expected cost per unit of time vector based on the operational phase (the system is operational).

$\mathbf{c}_d$: expected cost per unit of time vector based on the external damage phase (the system is operational).

$\mathbf{cr}_1$: expected corrective repair cost per unit of time vector based on the repair phase.

$\mathbf{cr}_2$: expected preventive maintenance cost per unit of time vector based on the preventive maintenance phase.

$H$: repairperson cost per unit of time (repairperson in the repair facility, working or not).

$F$: repairperson cost per unit of time while the repairperson is on vacation.

$C$: loss per unit of time during system downtime.

$G$: fixed cost each time that the repairperson returns (regardless of whether they stay or not).

*fcr*: fixed cost after a repairable failure.

*fmi*: fixed cost after a positive inspection (major inspection).

*fnu*: cost of a new unit (cost of a new system: $n{\cdot}fnu$).

When the system is in a specific state, it generates a corresponding net profit value. The system comprises both operational and non-operational units. To create the net profit vector, the costs and rewards linked to the online unit, as well as the expenses resulting from the repairperson's activities have been taken into account.

*Online unit*

If only the costs and rewards of the online unit are considered when the system is in the macro-state $\mathbf{E}_s^{k,v}$, a net reward for the phases within this macro-state is calculated. This results in the net profit vector specifically for the online unit when the repairperson is not present at their workplace. Then, for $k = 1,\ldots,n$,

$$\mathbf{nr}_s^{k,v} = \begin{cases} (B-F)\mathbf{e}_{mtd\varepsilon v} - \mathbf{c}_0 \otimes \mathbf{e}_{td\varepsilon v} - \mathbf{e}_{mt} \otimes \mathbf{c}_d \otimes \mathbf{e}_{\varepsilon v} & ; \quad s=0 \\ \begin{pmatrix} \mathbf{e}_{2^{s-1}} \otimes \left((B-F)\mathbf{e}_{mtd\varepsilon v} - \mathbf{c}_0 \otimes \mathbf{e}_{td\varepsilon v} - \mathbf{e}_{mt} \otimes \mathbf{c}_d \otimes \mathbf{e}_{\varepsilon v}\right) \\ \mathbf{e}_{2^{s-1}} \otimes \left((B-F)\mathbf{e}_{mtd\varepsilon v} - \mathbf{c}_0 \otimes \mathbf{e}_{td\varepsilon v} - \mathbf{e}_{mt} \otimes \mathbf{c}_d \otimes \mathbf{e}_{\varepsilon v}\right) \end{pmatrix} & ; \quad s=1,\ldots,k-1 \\ -(C+F)\mathbf{e}_{tv2^s} & ; \quad s=k. \end{cases}$$

Therefore, $\mathbf{nr}_s^{k,v}$ represents the net benefit vector generated by the online unit associated with the macro-state $\mathbf{E}_s^{k,v}$ at level 2, as described in Section 2.2. Thus, if *s*=0, he phases of this macro-state are $(k,0;i,j,h,u,w)$ with *i*=1, …, *m*; *j*=1, …, *t*; *h*=1, …, *d*; *u*=1, …, *ε* and *w*=1, …, *v*, yielding a total of *mtdεv* phases. The corresponding vector provides the net benefit for each of these phases, $(B-F)\mathbf{e}_{mtd\varepsilon v}-\mathbf{c}_0\otimes\mathbf{e}_{td\varepsilon v}-\mathbf{e}_{mt}\otimes\mathbf{c}_d\otimes\mathbf{e}_{\varepsilon v}$.

Let us now consider the case $0 < s < k$. The macro-state $\mathbf{E}_s^{k,v}$ is composed of the level 3 macro-states, $\mathbf{E}_{i_1,i_2,\ldots,i_s}^{k,v}$, where $i_j$ indicates whether the j-th unit requires corrective repair (=1) or preventive maintenance (=2) for *j*=1, 2, …, *s*.

Given that the repair technician is on vacation, there are no phases for corrective repair or maintenance time. Only the possible ordered combinations of the types of repairs to be performed ($i_1,\ldots, i_s$). should be considered. The net benefit associated with each of these combinations is $(B-F)\mathbf{e}_{mtd\varepsilon v}-\mathbf{c}_0\otimes\mathbf{e}_{td\varepsilon v}-\mathbf{e}_{mt}\otimes\mathbf{c}_d\otimes\mathbf{e}_{\varepsilon v}$.

The total number of arrangements of the types of repairs to be performed on the units in the repair facility facility is given by $2^s$. Therefore, the net profit vector for the corresponding phases is given by $\mathbf{e}_{2^s}\otimes\left((B-F)\mathbf{e}_{mtd\varepsilon v}-\mathbf{c}_0\otimes\mathbf{e}_{td\varepsilon v}-\mathbf{e}_{mt}\otimes\mathbf{c}_d\otimes\mathbf{e}_{\varepsilon v}\right)$.

Given that, in the case where the repairperson is not on vacation, the repair phases of the first unit -the one being serviced (corrective repair or preventive maintenance)- must be taken into account, the previous vector has been methodologically divided based on whether the first unit is undergoing corrective repair or preventive maintenance.

$$\mathbf{nr}_s^{k,v}=\mathbf{e}_{2^s}\otimes\left((B-F)\mathbf{e}_{mtd\varepsilon v}-\mathbf{c}_0\otimes\mathbf{e}_{td\varepsilon v}-\mathbf{e}_{mt}\otimes\mathbf{c}_d\otimes\mathbf{e}_{\varepsilon v}\right)$$
$$=\begin{pmatrix}\mathbf{e}_{2^{s-1}}\otimes\left((B-F)\mathbf{e}_{mtd\varepsilon v}-\mathbf{c}_0\otimes\mathbf{e}_{td\varepsilon v}-\mathbf{e}_{mt}\otimes\mathbf{c}_d\otimes\mathbf{e}_{\varepsilon v}\right)\\ \mathbf{e}_{2^{s-1}}\otimes\left((B-F)\mathbf{e}_{mtd\varepsilon v}-\mathbf{c}_0\otimes\mathbf{e}_{td\varepsilon v}-\mathbf{e}_{mt}\otimes\mathbf{c}_d\otimes\mathbf{e}_{\varepsilon v}\right)\end{pmatrix}.$$

Finally, we consider the case $\mathbf{nr}_k^{k,v}$. Following the previous reasoning, for each arrangement of the units in the repair channel (all broken), the net profit is given by the losses due to the lack of operational units and the vacation cost. Thus, for each phase (phase of the external shock time and vacation time), there is a cost of $-(C+F)\mathbf{e}_{tv}$. Since the possible number of arrangements in the repair facility based on the type of task to be performed by the technician is equal to $2^k$, we have

$$\mathbf{nr}_k^{k,v}=-(C+F)\mathbf{e}_{tv2^s}.$$

If the repairperson is not on vacation, the reasoning is analogous to the previous case, except that the corrective repair or preventive maintenance phases must be considered when there are units in the repair facility. Thus, we have that

$$\mathbf{nr}_s^{k,nv}=\begin{cases}(B-H)\mathbf{e}_{mtd\varepsilon}-\mathbf{c}_0\otimes\mathbf{e}_{td\varepsilon}-\mathbf{e}_{mt}\otimes\mathbf{c}_d\otimes\mathbf{e}_{\varepsilon} & ;\quad s=0\\ \begin{pmatrix}\mathbf{e}_{2^{s-1}}\otimes\left((B-H)\mathbf{e}_{mtd\varepsilon z_1}-\mathbf{c}_0\otimes\mathbf{e}_{td\varepsilon z_1}-\mathbf{e}_{mt}\otimes\mathbf{c}_d\otimes\mathbf{e}_{\varepsilon z_1}\right)\\ \mathbf{e}_{2^{s-1}}\otimes\left((B-H)\mathbf{e}_{mtd\varepsilon z_2}-\mathbf{c}_0\otimes\mathbf{e}_{td\varepsilon z_2}-\mathbf{e}_{mt}\otimes\mathbf{c}_d\otimes\mathbf{e}_{\varepsilon z_2}\right)\end{pmatrix} & ;\quad s=1,\ldots,k-1\\ \begin{pmatrix}-\mathbf{e}_{2^{s-1}}\otimes(C+H)\mathbf{e}_{tz_1}\\ -\mathbf{e}_{2^{s-1}}\otimes(C+H)\mathbf{e}_{tz_2}\end{pmatrix} & ;\quad s=k.\end{cases}$$

The vector $\mathbf{nr}^{k}$ for the macro-state $\mathbf{U}^k$, $k$ units in the system, is given, for $N=k-R+1$, by

$$\mathbf{nr}^k=\begin{pmatrix}\mathbf{nr}_0^{k,v}\\ \mathbf{nr}_1^{k,v}\\ \vdots\\ \mathbf{nr}_k^{k,v}\\ \mathbf{nr}_N^{k,nv}\\ \vdots\\ \mathbf{nr}_k^{k,nv}\end{pmatrix}\text{ for } k\le R \text{ and } \mathbf{nr}^k=\begin{pmatrix}\mathbf{nr}_0^{k,nv}\\ \mathbf{nr}_1^{k,nv}\\ \vdots\\ \mathbf{nr}_k^{k,nv}\end{pmatrix}\text{ for } k>R.$$

Finally, the net reward vector **nr** (only costs and rewards from the online unit) is given by $\mathbf{nr}=\left(\mathbf{nr}^n,\mathbf{nr}^{n-1},\ldots,\mathbf{nr}^1\right)'$.

*Cost from the repair facility*

The net reward is completed with the cost associated with corrective repair and preventive maintenance. Costs are only possible when there are units in the repair facility, and the repairer is at the workplace. Therefore, for the macro-state $\mathbf{E}_s^{k,v}$, this vector is $\mathbf{nc}_s^{k,v}=\mathbf{0}_{tv2^s(md\varepsilon)^{I\{s<k\}}}$ and it is for the macro-state $\mathbf{E}_s^{k,nv}$,

$$\mathbf{nc}_s^{k,nv}=\begin{pmatrix}\mathbf{e}_{2^{s-1}}\otimes\mathbf{e}_{t(md\varepsilon)^{I\{s<k\}}}\otimes\mathbf{cr}_1\\ \mathbf{e}_{2^{s-1}}\otimes\mathbf{e}_{t(md\varepsilon)^{I\{s<k\}}}\otimes\mathbf{cr}_2\end{pmatrix}\quad;\quad s=1,\ldots,k\,.$$

This vector contains the cost associated with repair or preventive maintenance for each phase of the macro-state $E_s^{k,nv}$. If $s<k$, there is one unit online, and therefore, all phases of the main unit's operating time, the time until an external shock, the damage states due to the external shock, and the time until inspection must be considered. This results in a total of mtdε fases, ($\mathbf{e}_{mtd\varepsilon}$).

In the repair facility, there are $s$ units, the one being repaired and $s-1$ in the repair queue. The number of distinct arrangements of the types of repairs in the queue is $2^{s-1}$, ($\mathbf{e}_{2^{s-1}}$).

Finally, the cost vector when a unit is undergoing corrective repair, according to the repair phases, is given by **cr**1. Therefore, in this case, we have $\mathbf{e}_{2^{s-1}} \otimes \mathbf{e}_{mtd\varepsilon} \otimes \mathbf{cr}_1$. Analogously, for the case where the unit being serviced is undergoing preventive maintenance, we have $\mathbf{e}_{2^{s-1}} \otimes \mathbf{e}_{mtd\varepsilon} \otimes \mathbf{cr}_2$.

What happens when $s = k$? In this case, there are no units in the online position, so only the phases of the time until external shock and the possible arrangements of the units in the repair queue need to be considered. If the unit being serviced by the repairperson is undergoing corrective repair, we have $\mathbf{e}_{2^{s-1}} \otimes \mathbf{e}_t \otimes \mathbf{cr}_1$ and if it is undergoing preventive maintenance, we have $\mathbf{e}_{2^{s-1}} \otimes \mathbf{e}_t \otimes \mathbf{cr}_2$.

Analogously to the case above, the macro-state $\mathbf{U}^k$ have been built from these vectors.

It is, for $k \leq R$, $\mathbf{nc}^k = \begin{pmatrix} \mathbf{nc}_0^{k,nv} \\ \mathbf{nc}_1^{k,nv} \\ \vdots \\ \mathbf{nc}_k^{k,nv} \end{pmatrix}$ and for $k > R$, $\mathbf{nc}^k = \begin{pmatrix} \mathbf{nc}_0^{k,v} \\ \mathbf{nc}_1^{k,v} \\ \vdots \\ \mathbf{nc}_k^{k,v} \\ \mathbf{nc}_N^{k,nv} \\ \vdots \\ \mathbf{nc}_k^{k,nv} \end{pmatrix}$, where for $N = k–R+1$.

The cost vector associated to the maintenance facility, **nc,** is given by $\mathbf{nc} = \begin{pmatrix} \mathbf{nc}^n \\ \mathbf{nc}^{n-1} \\ \vdots \\ \mathbf{nc}^1 \end{pmatrix}$.

From both vectors, **nr** and **nc** the net reward/cost vector is obtained, **c** =**nr**–**nc**.

In Appendix C, an example is provided to demonstrate the methodology.

### 4.4.2. Mean net total profit

It is of great interest to have measures that allow the comparison of models based on net profit. Therefore, total net profit is constructed under transient and steady-state regime. This measure takes into account the evolving benefits/costs and fixed costs incurred by different events.

The mean net profit up to time $t$ generated by the main unit is given by,

$$\Phi_w(t) = \int_0^t \mathbf{p}(t)\,dt \cdot \mathbf{nr}\,. \quad (6)$$

On the other hand, the mean costs up to time $t$ incurred in the repair channel, whether through corrective repair or preventive maintenance, can be found by

$$\Phi_{rf}(t) = \int_0^t \mathbf{p}(t)\,dt \cdot \mathbf{nc}\,. \quad (7)$$

In stationary regime, these measures can be interpreted as the mean net profit/cost per unit of time. They respectively have the expression from (6) and (7), $\Phi_w = \boldsymbol{\pi} \cdot \mathbf{nr}$ and $\Phi_{rf} = \boldsymbol{\pi} \cdot \mathbf{nc}$.

If the fixed costs are incorporated into the previous measures, the mean net total profit up to time *t* is defined from (6), (7) and the measures defined in Table 1 as,

$$\begin{aligned}\Phi(t) = \Phi_w(t) - \Phi_{rf}(t) - \left(1 + \Lambda^{NS}(t)\right) \cdot n \cdot fnu - \Lambda^{rep}(t) \cdot fcr \\ - \Lambda^{mi}(t) \cdot fmi - \left(\Lambda^{ret}(t) + \Lambda^{ret-be}(t)\right) \cdot G.\end{aligned} \tag{8}$$

Finally, the mean net total profit per unit of time in stationary regime from (9) is

$$\Phi = \Phi_w - \Phi_{rf} - \Lambda^{NS} \cdot n \cdot fnu - \Lambda^{rep} \cdot fcr - \Lambda^{mi} \cdot fmi - \left(\Lambda^{ret} + \Lambda^{ret-be}\right) \cdot G. \tag{9}$$

## 5. A Numerical Example

Utilizing the suggested methodology allows for the modeling of various real-world systems, including backup generators in enterprises, spare hardware in IT infrastructures, or power generators in civil engineering systems. Clearly, there are associated costs with operation of a complex system that it is desirable to optimize. When examining a system like the one developed in this study, delving into the optimal number of units in cold standby becomes particularly intriguing, along with assessing the profitability of preventive maintenance. Additionally, it is essential to identify the optimal distribution of vacation time and determine the corresponding R-value of the proposed vacation policy. This analysis has the potential to furnish valuable information for making well-informed decisions in practical applications.

### 5.1. The system

Multiple cold standby systems with an identical behavior of the online unit and repair facility to identify the optimal configuration are assumed. Systems with 2, 3, and 4 units initially, both with and without preventive maintenance, are taken into account. Furthermore, a vacation policy is across all systems, considering exponential and generalized Erlang distributions (both phase-type distributions with representations given in Table 2) for random vacation time, encompassing all possible values of *R*.

The behavior is as follows when each unit assumes the online position. The online unit comprises four internal performance states, with the initial two regarded as minor damage states and the subsequent two categorized as major damage states. Two kinds of transitions are possible, to the next damage state or to repairable or non-repairable failure. The transitions rates to repairable failure are four times higher than the rates to non-repairable failure, contingent upon the performance state. The mean time up to internal failure is 45.333 units of time.

The online system is susceptible to external shocks. The intervals between two successive external disturbances are governed by a phase-type distribution with meat time 11.2 units of time.

When the system experiences an external shock, it can be produced by three possible consequences,

- Total failure with a probability equal to 0.2.
- Internal performance aggravation of the online unit (probability of aggravation transition matrix denoted as **W**), even potentially leading to repairable or non-repairable failure from different phases of internal operation. The probabilities for these transitions are specified in the vectors $\mathbf{W}_r^0$ and $\mathbf{W}_{nr}^0$, respectively.
- Cumulative damage leading to non-repairable failure is considered. The second external shock always results in non-repairable failure, governed by the initial vector $\boldsymbol{\omega}$ and matrix $\mathbf{D}$.

Inspections occur at random intervals, with a mean time of 16.667 units between two consecutive inspections when the online position is occupied. The online unit is sent to preventive maintenance when certain conditions are met, such as observing any of the last two internal phases or experiencing that an external shock has already occurred.

Table 2 shows the representation of the phase type distributions associated to the times embedded in the system and Table 3 the auxiliary matrices.

To optimize the system, costs and rewards are taken into consideration. A gross profit of $B$=70 when the system is operational is assumed, and this also represents the loss per unit of time when the system is not operational, $C$ =70. The online unit incurs operational costs that vary depending on the operational phase. These costs are represented by the vector $\mathbf{c}_0$ = (6, 14, 32, 42)'.

The repairperson can either be on vacation or at its workplace. If this is on his workplace a cost per unit of time equal to $H$ = 20 is produced, whereas if he is on vacation it is $F$ = 4. A cost equal to $G$ = 5 is fixed for each time that the repairperson returns from a vacation period.

<table>
<tr><th>Internal time</th><th>External shock time</th><th>Inspection time</th></tr>
<tr><td>

$\boldsymbol{\alpha} = (1,0,0,0)$

$\mathbf{T} = \begin{pmatrix} -0.04 & 0.02 & 0 & 0 \\ 0 & -0.03 & 0.02 & 0 \\ 0 & 0 & -0.1 & 0.04 \\ 0 & 0 & 0 & -0.4 \end{pmatrix}$

$\mathbf{T}_r^0 = \begin{pmatrix} 0.016 \\ 0.008 \\ 0.048 \\ 0.32 \end{pmatrix} \quad \mathbf{T}_{nr}^0 = \begin{pmatrix} 0.004 \\ 0.002 \\ 0.012 \\ 0.08 \end{pmatrix}$

Mean time: 45.3333 u.t.

</td><td>

$\boldsymbol{\gamma} = (1,0)$

$\mathbf{L} = \begin{pmatrix} -0.1 & 0.06 \\ 0 & -0.5 \end{pmatrix}$

Mean Time: 11.2 u.t.

</td><td>

$\boldsymbol{\eta} = (1,0)$

$\mathbf{M} = \begin{pmatrix} -0.2 & 0.15 \\ 0.5 & -0.6 \end{pmatrix}$

Mean time: 16.6667 u.t.

</td></tr>
<tr><th>Corrective repair time</th><th>Preventive maintenance time</th><th>Vacation time</th></tr>
<tr><td>

$\boldsymbol{\beta}_1 = (1,0,0)$

$\mathbf{S}_1 = \begin{pmatrix} -0.8 & 0.5 & 0.2 \\ 0.3 & -0.8 & 0.4 \\ 0.4 & 0.1 & -0.7 \end{pmatrix}$

Mean corrective repair time: 7.7640 u.t.

</td><td>

$\boldsymbol{\beta}_2 = (1,0,0)$

$\mathbf{S}_2 = \begin{pmatrix} -0.8 & 0.2 & 0.05 \\ 0.05 & -0.9 & 0.2 \\ 0.1 & 0.1 & -0.8 \end{pmatrix}$

Mean corrective repair time: 1.7487 u.t.

</td><td>

Exponential distribution

$\upsilon = 1$<br>
$\mathbf{V} = -a$

Generalized Erlang distribution

$\upsilon = (1,0)$

$\mathbf{V} = \begin{pmatrix} -a & a \\ 0 & -b \end{pmatrix}$

</td></tr>
</table>

Table 2. Phase-type distributions of the times embedded in the system

| Aggravation internal probability matrix after external shock | Cumulative damage matrix | Matrices to determine minor or major inspection |
|---|---|---|
| $\mathbf{W}=\begin{pmatrix}0.1 & 0.05 & 0.2 & 0.05\\ 0 & 0.05 & 0.2 & 0.05\\ 0 & 0 & 0.2 & 0.05\\ 0 & 0 & 0 & 0.05\end{pmatrix}$ <br> $\mathbf{W}_r^0=\begin{pmatrix}0.6\\ 0.6\\ 0.65\\ 0.65\end{pmatrix}$ $\mathbf{W}_{nr}^0=\begin{pmatrix}0\\ 0.1\\ 0.1\\ 0.3\end{pmatrix}$ | $\boldsymbol{\omega}=(1,0)$ <br> $\mathbf{D}=\begin{pmatrix}0 & 1\\ 0 & 0\end{pmatrix}$ | Minor Internal damage $\mathbf{U}_1=\begin{pmatrix}1 & 0 & 0 & 0\\ 0 & 1 & 0 & 0\\ 0 & 0 & 0 & 0\\ 0 & 0 & 0 & 0\end{pmatrix}$ Minor Internal damage $\mathbf{U}_2=\begin{pmatrix}0 & 0 & 0 & 0\\ 0 & 0 & 0 & 0\\ 0 & 0 & 1 & 0\\ 0 & 0 & 0 & 1\end{pmatrix}$ <br> Minor External damage $\mathbf{V}_1=\begin{pmatrix}1 & 0\\ 0 & 0\end{pmatrix}$ Minor External damage $\mathbf{V}_2=\begin{pmatrix}0 & 0\\ 0 & 1\end{pmatrix}$ |

Table 3. Complementary matrices of the system

In the event of a repairable failure of the online unit, it is sent to the repair facility for corrective repair. A fixed cost of *fcr* = 10 is associated with each repairable failure, and once in corrective repair, the cost is determined by the corresponding state, and is denoted as $\mathbf{cr}_1$ = (20, 20, 20)'. Analogously it occurs for major damage and preventive maintenance. If an inspection identifies major damage on the main unit, it is sent to the repair facility for preventive maintenance, incurring a fixed cost of *fmi* = 4. Once in preventive maintenance, the cost associated to each stage is given by $\mathbf{cr}_2$ = (10, 10, 10)'.

When an external shock occurs, an external damage is produced. Two stages for external damage are considered in this example. The costs associated to these stages while the system is operational is given by $\mathbf{c}_d$ = (1, 2)'.

The cost of a new system, after non-repairable failure of all units, is *fnu* = 150.

### 5.2. Optimization

By considering the aforementioned behavior of the online unit, we have optimized the distribution of vacation time for systems with $n$ = 2, 3, 4 units, and $R$ = 1,..., $n$, considering both exponential and generalized Erlang distribution classes with 2 stages. This optimization applies to scenarios with and without preventive maintenance. To achieve this, the mean net total profit per unit of time in the stationary regime has been maximised, as outlined in (9) in Section 4.4.2.

Thus, for each of the 36 distinct models, the function net total profit per unit of time in the stationary regime is optimized, where the parameters involved are $\mathbf{x}$=$a$ or $\mathbf{x}$=($a$, $b$). These parameters correspond to the phase-type distributions of vacation time, exponential, and Erlang, respectively. Both distributions are phase-type, and their representation is presented in Table 2.

Therefore, in this case, the function $\Phi$ depends on the parameter vector $\mathbf{x}$ and must be maximized for each model. Then,

$$\mathbf{x} \text{ s.t. } \max_{\mathbf{x}>0} \Phi(\mathbf{x}).$$

The optimization was carried out using the Matlab program, employing the gradient-based Interior-Point method. This method can optimally handle large and complex problems.

Figure 2 illustrates the 36 models optimized based on vacation time, and Table 4 presents the corresponding values. It can be observed that the optimal system is achieved in the case of 4 units and R=3, where the vacation time follows a generalized Erlang distribution with PH representation

$$\upsilon = (1,0), \quad \mathbf{V} = \begin{pmatrix} -0.8297104 & 0.8297104 \\ 0 & -0.8297099 \end{pmatrix},$$

and with preventive maintenance. The optimum mean net total profit per unit of time is 8.2506. It is noticeable that, depending on the systems, losses may occur.

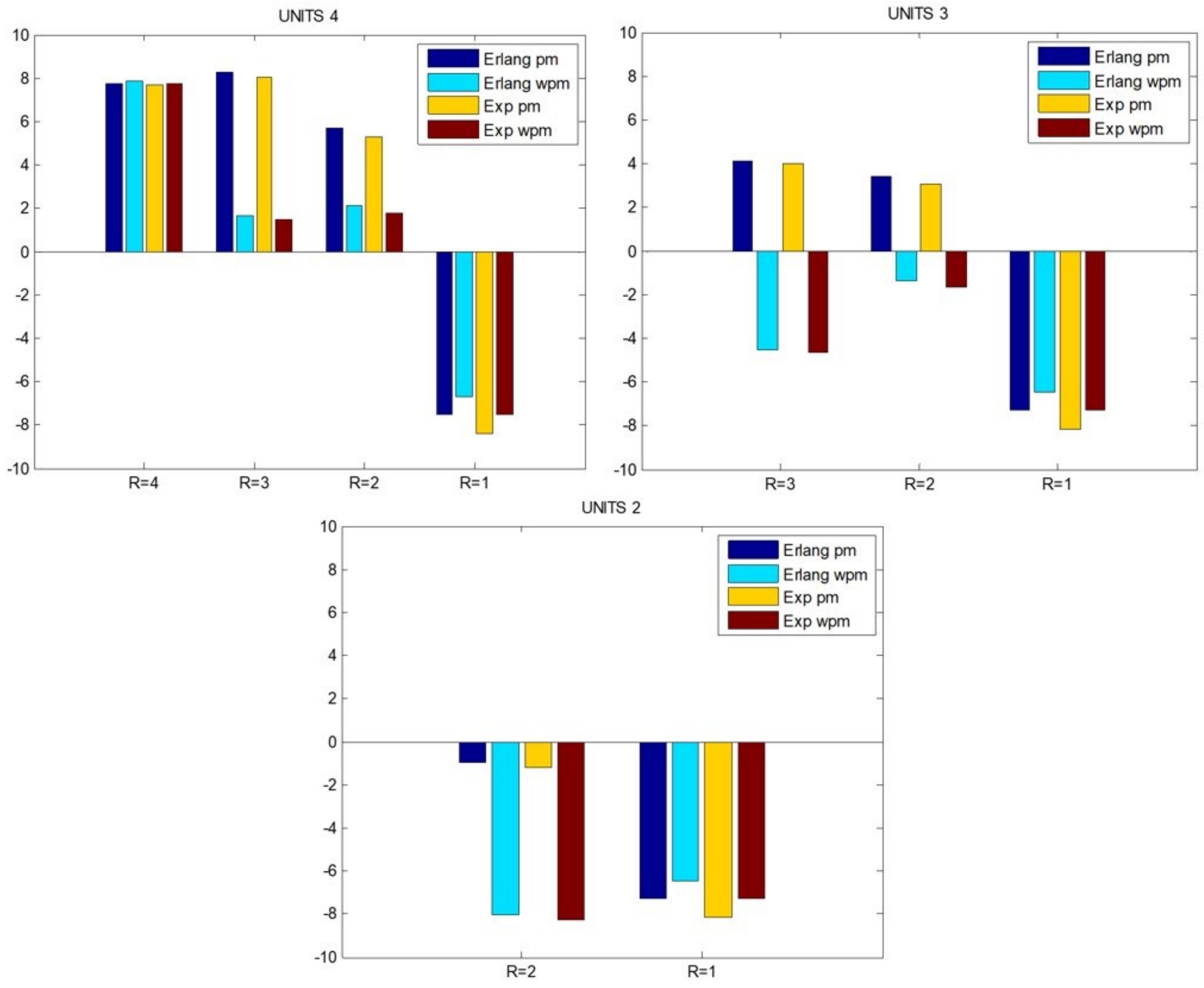

Figure 2. Optimal mean net total profit in stationary regime for the different systems

| | | *n*=4 | | | | *n*=3 | | | *n*=2 | |
|---|---|---|---|---|---|---|---|---|---|---|
| | | *R=4* | *R=3* | *R=2* | *R=1* | *R=3* | *R=2* | *R=1* | *R=2* | *R=1* |
| **Erlang** | PM | 7.7716 | **8.2506** | 5.6873 | −7.5407 | 4.1268 | 3.3951 | −7.3085 | −0.9709 | −7.2861 |
| | Without PM | 7.8445 | 1.6722 | 2.1366 | −6.6852 | −4.5597 | −1.3608 | −6.4767 | −8.0367 | −6.4519 |
| **Exp** | PM | 7.7012 | 8.0610 | 5.2838 | −8.4092 | 4.0037 | 3.0411 | −8.1811 | −1.2277 | −8.1638 |
| | Without PM | 7.7758 | 1.4974 | 1.7571 | −7.5106 | −4.6697 | −1.6927 | −7.3063 | −8.2750 | −7.2873 |

Table 4. Optimal mean net total profit for the systems.

After optimizing the systems, it becomes crucial to assess their performance. To achieve this, the stationary availability, described in (2), for each system is calculated and conducted a comparative analysis. This is illustrated in Figure 3 and Table 5. The systems that yield the optimal mean net profit do not necessarily coincide with those that offer the highest availability. In this scenario, the optimal availability is achieved for the case $k$ = 4 and R = 4, with no preventive maintenance, and under an exponentially distributed vacation time. This outcome might be typical, as more time spent at the workplace generally leads to higher availability but may result in a lower mean net profit.

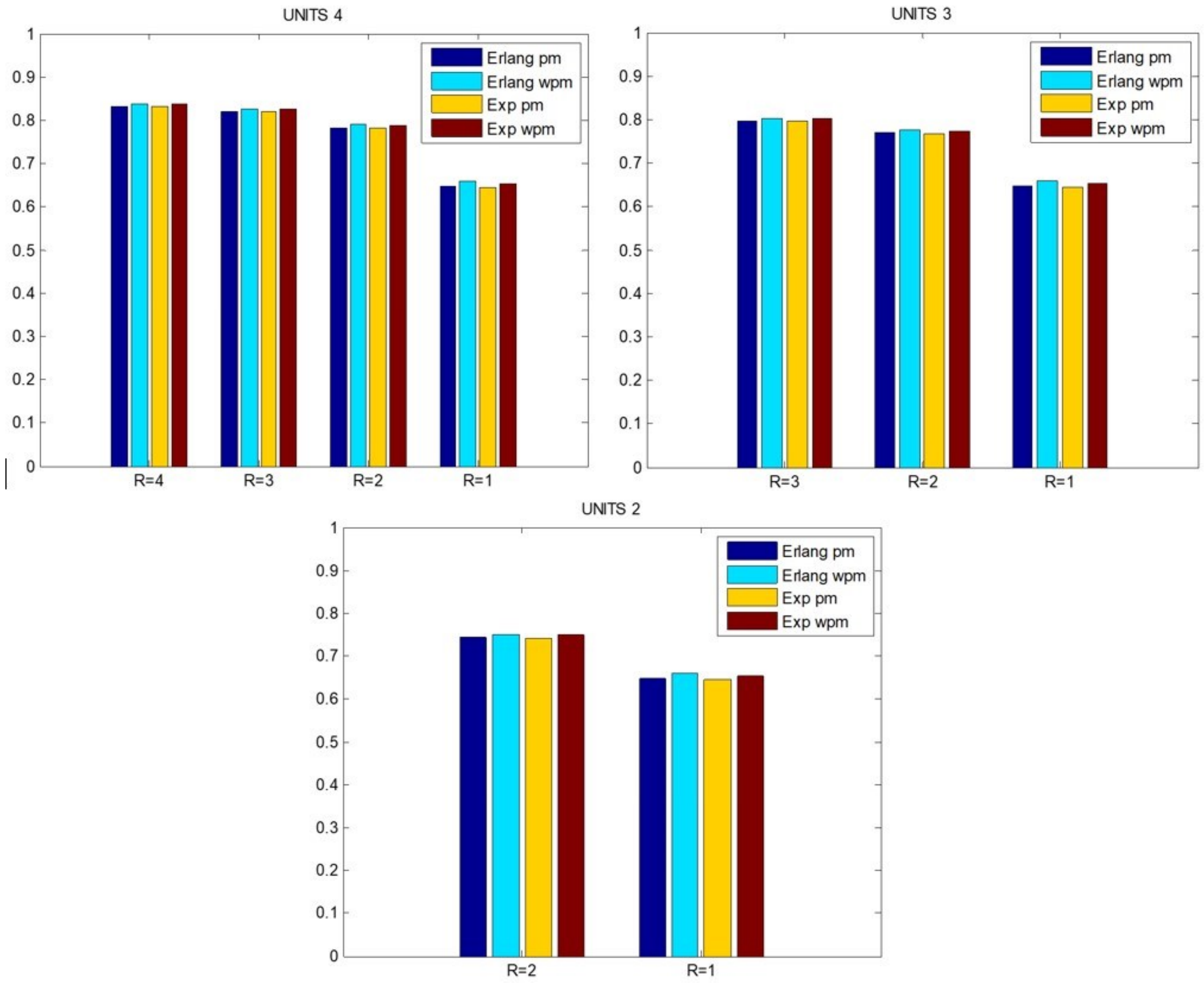

Figure 3. Optimal stationary availability for the different systems

| | | $n$=4 | | | | $n$=3 | | | $n$=2 | |
|---|---|---|---|---|---|---|---|---|---|---|
| | | $R$=4 | $R$=3 | $R$=2 | $R$=1 | $R$=3 | $R$=2 | $R$=1 | $R$=2 | $R$=1 |
| **Erlang** | PM | 0.8323 | 0.8210 | 0.7832 | 0.6481 | 0.7967 | 0.7695 | 0.6482 | 0.7435 | 0.6482 |
| | Without PM | **0.8384** | 0.8269 | 0.7900 | 0.6581 | 0.8030 | 0.7764 | 0.6581 | 0.7508 | 0.6582 |
| **Exp** | PM | 0.8320 | 0.8202 | 0.7814 | 0.6438 | 0.7961 | 0.7679 | 0.6439 | 0.7424 | 0.6440 |
| | Without PM | 0.8380 | 0.8260 | 0.7881 | 0.6540 | 0.8024 | 0.7748 | 0.6541 | 0.7496 | 0.6542 |

Table 5. Optimal stationary availability values for the systems.

## 5.3 The optimal system. Measures

From an economic standpoint, it has been determined that the optimal system consists of four units, with vacation time period following an Erlang distribution, and the repairer remains in the repair channel after vacations if fewer than 3 operational units are observed when he returns ($R$=3). Several associated measures for this system have been obtained as described in Section 4. Table 6 shows the rate of being in the different macro-states, as it has been developed in (4), and 7 display the availability, (2), and some measures introduced in Table 1 under steady-state conditions.

| $\Psi^{v}_{k,s}$ | | | | | |
|---|---|---|---|---|---|
| | $s=0$ | $s=1$ | $s=2$ | $s=3$ | $s=4$ |
| $k=4$ | 0.0677 | 0.0847 | 0.0096 | 0.0008 | 0.0001 |
| $k=3$ | 0.0963 | 0.0159 | 0.0015 | 0.0001 | |
| $\Psi^{nv}_{k,s}$ | | | | | |
| | $s=0$ | $s=1$ | $s=2$ | $s=3$ | $s=4$ |
| $k=4$ | | | 0.0301 | 0.0149 | 0.0071 |
| $k=3$ | | 0.0566 | 0.0364 | 0.0203 | |
| $k=2$ | 0.1330 | 0.0724 | 0.0402 | | |
| $k=1$ | 0.2012 | 0.1112 | | | |

Table 6. Time rate in macro-state $E^{v}_{k,s}$ and $E^{nv}_{k,s}$ in stationary regime

| | |
|---|---|
| **Availability** $(A)$ | 0.8210 |
| **Average proportion of repairable failures** $(\Lambda^{rep})$ | 0.0487 |
| **Average proportion of major inspections** $(\Lambda^{mi})$ | 0.0064 |
| **Average proportion of non-repairable failures** $(\Lambda^{nr})$ | 0.0261 |
| **Average proportion of returns to work** $(\Lambda^{ret})$ | 0.0191 |
| **Average proportion of returns to work and start of a new vacation period** $(\Lambda^{ret-be})$ | 0.0985 |
| **Average proportion of vacation periods after repair** $(\Lambda^{after})$ | 0.0125 |
| **Average proportion of new systems** ( $\Lambda^{NS}$ ) | 0.0065 |

Table 7. Proportional number of events in stationary regime

## 6. Conclusions

In this work, a complex redundant reliability system with loss of units and a vacation policy is modelled using a MMAP in continuous time. This modelling is carried out algorithmically and matrix-based. For its development, phase-type distributions have been considered for the embedded times in the model, a fact that would allow considering it as a general model since this type of distributions fulfils that it is dense in the set of non-negative probability distributions.

The proposed modelling allows for a more accessible interpretation of results, as well as their computational implementation. The main unit is subject to multiple events to encompass applicability in case they exist, and a restriction can be easily applied in such a case. Failures due to wear, whether repairable or not, external shocks with multiple consequences, and random inspection introducing the study of preventive maintenance are introduced.

The constructed MMAP has allowed the creation of performance measures of interest, as well as the analysis of different events always in a matrix-algorithmic form. The operation of a complex system is inevitably linked to the economic aspect. Therefore, costs and rewards have been introduced based on the state of the overall system depending on multiple factors. This has allowed the construction of new net profit measures.

To highlight the versatility of the model, an optimization problem has been carried out. Multiple systems with identical characteristics to the online unit have been considered, and the average net profit has been optimized, obtaining the optimal model according to vacation policy, vacation time distribution, and the incorporation of preventive maintenance.

This work highlights the value of the developed methodology. This methodology enables the algorithmic modeling of various complex systems in the field of reliability, such as unitary systems, systems with standby units, or $k$-out-of-$n$: $G$ systems. An interesting first generalization of the proposed work is the algorithmic modeling of a redundant system with an indeterminate number of online units. On the other hand, in the proposed system, the inspection time is considered negligible, as may be the case with integrated sensors. However, there may be systems where this assumption does not match their behavior, requiring the introduction of a random inspection time. In both of the aforementioned cases, the developed methodology is applicable by modifying and expanding the corresponding state space as well as the matrix structure.

The entire development has been carried out in transient and stationary regime, obtaining the steady-state distribution of the model using analytic-matrix methods. Both the modeling and the measures and results have been computationally implemented with Matlab.

## Acknowledgements

This paper is partially supported by the project FQM-307 of the Government of Andalusia (Spain) and by the project PID2023-149087NB-I00 of the Spanish Ministry of Science and Innovation (also supported by the FEDER programme). Additionally, the first author would like to express their gratitude for financial support by the IMAG–María de Maeztu grant CEX2020-001105-M/AEI/10.13039/501100011033.

Declaration of generative AI and AI-assisted technologies in the writing process

During the preparation of this work the authors used webpage chat.openai.com in order to improve language and readability, with caution. After using this webpage, the authors reviewed and edited the content as needed and take full responsibility for the content of the publication.

*Appendix A*

Matrix $\mathbf{H}_A$

This matrix block contains the transition intensities from the operational phases of the online unit to a repairable failure.

The repairable failure can be triggered by the following events:

- Internal repairable failure without external shock: This occurs via the column vector $\mathbf{T}_r^0$, after which a new unit occupies the online position with an initial distribution $\boldsymbol{\alpha}$ for internal performance. Since only one transition can occur at any given time, a external shock is not present, the new online unit begins accumulating external damage according to matrix $\mathbf{e\omega}$ and a new inspection period starts $\mathbf{e\eta}$. Then,

$$\mathbf{T}_r^0\boldsymbol{\alpha}\otimes\mathbf{I}\otimes\mathbf{e\omega}\otimes\mathbf{e\eta}.$$

- External shock and repairable failure from modification of internal behavior: An external shock occurs without leading to a catastrophic failure, and a new external shock period begins, $\mathbf{L}^0\boldsymbol{\gamma}\left(1-\omega^0\right)$. This external shock modifies the internal structure of the online unit, resulting in a repairable failure, and the new online unit is reinitialized, $\mathbf{W}_r^0\boldsymbol{\alpha}$. Finally, the external shock does not cause a non-repairable failure due to cumulative external damage, and this cumulative external damage is reset for the new online unit, $\mathbf{De\omega}$. A new inspection period starts $\mathbf{e\eta}$. Then,

$$\mathbf{W}_r^0\boldsymbol{\alpha}\otimes\mathbf{L}^0\boldsymbol{\gamma}\left(1-\omega^0\right)\otimes\mathbf{De\omega}\otimes\mathbf{e\eta}.$$

Therefore, we have $\mathbf{H}_A=\left[\mathbf{T}_r^0\boldsymbol{\alpha}\otimes\mathbf{I}\otimes\mathbf{e\omega}+\mathbf{W}_r^0\boldsymbol{\alpha}\otimes\mathbf{L}^0\boldsymbol{\gamma}\left(1-\omega^0\right)\otimes\mathbf{De\omega}\right]\otimes\mathbf{e\eta}$.

If, after the repairable failure, no reserve units are available, then

$$\mathbf{H}_A^{'}=\left[\mathbf{T}_r^0\otimes\mathbf{I}\otimes\mathbf{e}+\mathbf{W}_r^0\otimes\mathbf{L}^0\boldsymbol{\gamma}\left(1-\omega^0\right)\otimes\mathbf{De}\right]\otimes\mathbf{e}.$$

Matrix $\mathbf{H}_0$

Matrix $\mathbf{H}_0$ contains the transition intensities when they do not result in any event *A*, *B*, or *C*. This can occur in five different situations:

- Only a transition in the internal behavior of the main unit (**T**). In this case, we have $\mathbf{T}\otimes\mathbf{I}\otimes\mathbf{I}\otimes\mathbf{I}$.
- Only a transition in the time until external shock occurs without it actually happening. We have $\mathbf{I}\otimes\mathbf{L}\otimes\mathbf{I}\otimes\mathbf{I}$.

  Clearly, these first two cases can be expressed as:

$$\mathbf{T}\otimes\mathbf{I}\otimes\mathbf{I}\otimes\mathbf{I}+\mathbf{I}\otimes\mathbf{L}\otimes\mathbf{I}\otimes\mathbf{I}=\left(\mathbf{T}\otimes\mathbf{I}+\mathbf{I}\otimes\mathbf{L}\right)\otimes\mathbf{I}\otimes\mathbf{I}=\mathbf{T}\oplus\mathbf{L}\otimes\mathbf{I}\otimes\mathbf{I},$$

  where $\oplus$ is the Kronecker sum.

- Only a transition in the inspection time without it occurring. This results in

$$\mathbf{I}\otimes\mathbf{I}\otimes\mathbf{I}\otimes\mathbf{M}.$$

- An external shock occurs without causing a total direct failure ($\mathbf{L}^0\boldsymbol{\gamma}\left(1-\omega^0\right)$), but it may modify the internal behavior and external damage, without causing a failure, through the probability matrices respectively **W** and **D**. The inspection time remains unchanged. The matrix is given as

$$\mathbf{W}\otimes\mathbf{L}^0\boldsymbol{\gamma}\left(1-\omega^0\right)\otimes\mathbf{D}\otimes\mathbf{I}.$$

- Transition in the inspection time resulting in an inspection ($\mathbf{M}^0\boldsymbol{\eta}$) that observes minor internal and external damage. In this case, we have

$$\mathbf{U}_1\otimes\mathbf{I}\otimes\mathbf{V}_1\otimes\mathbf{M}^0\boldsymbol{\eta}.$$

The final transition matrix is given as

$$\mathbf{H}_0=\mathbf{T}\oplus\mathbf{L}\otimes\mathbf{I}\otimes\mathbf{I}+\mathbf{I}\otimes\mathbf{I}\otimes\mathbf{I}\otimes\mathbf{M}+\mathbf{W}\otimes\mathbf{L}^0\boldsymbol{\gamma}\left(1-\omega^0\right)\otimes\mathbf{D}\otimes\mathbf{I}+\mathbf{U}_1\otimes\mathbf{I}\otimes\mathbf{V}_1\otimes\mathbf{M}^0\boldsymbol{\eta}.$$

*Appendix B*

*The matrices* $\mathbf{D}^A$ *and* $\mathbf{D}^B$

The matrices $\mathbf{D}^A$ and $\mathbf{D}^B$ consist of matrix blocks that encompass the transitions between macro-states $\mathbf{U}^k$. Obviously, they are diagonal matrix blocks.

$\mathbf{D}_k^Y$ : transition intensities for scenarios where: *k* units, event *Y* (*A* or *B*). This applies for values of *k*=1, ..., *n*.

$$
\mathbf{D}^{Y}=\begin{pmatrix}
\mathbf{D}_{n}^{Y} & & & & \\
 & \mathbf{D}_{n-1}^{Y} & & & \\
 & & \mathbf{D}_{n-2}^{Y} & & \\
 & & & \ddots & \\
 & & & & \mathbf{D}_{1}^{Y}
\end{pmatrix}.
$$

- If $k \leq R-1$ (no vacations for the repairperson)

$$
\mathbf{D}_{k}^{Y}=\begin{array}{c}
E_{0}^{k,nv}\\ E_{1}^{k,nv}\\ \vdots \\ E_{k-1}^{k,vn}\\ E_{k}^{k,nv}
\end{array}
\overset{\begin{array}{ccccc} E_{0}^{k,nv} & E_{1}^{k,nv} & E_{2}^{k,nv} & E_{k-1}^{k,nv} & E_{k}^{k,nv}\end{array}}{\begin{pmatrix}
\mathbf{0} & \mathbf{D}_{01}^{Y,k,nv} & & & \\
 & \mathbf{0} & \mathbf{D}_{12}^{Y,k,nv} & & \\
 & & \ddots & \ddots & \\
 & & & \mathbf{0} & \mathbf{D}_{k-1,k}^{Y,k,nv}\\
 & & & & \mathbf{0}
\end{pmatrix}}
$$

- If $k \geq R$ (vacations or not for the repairperson)

$$
\mathbf{D}_{k}^{Y}=
\begin{array}{c|cccccccccc|ccccc}
 & E_{0}^{k,v} & E_{1}^{k,v} & \cdots & E_{N-1}^{k,v} & E_{N}^{k,v} & E_{N+1}^{k,v} & \cdots & E_{k-1}^{k,v} & E_{k}^{k,v} & & E_{N}^{k,nv} & E_{N+1}^{k,nv} & E_{N+2}^{k,nv} & E_{k-1}^{k,nv} & E_{k}^{k,nv}\\
\hline
E_{0}^{k,v} & \mathbf{0} & \mathbf{D}_{0,1}^{Y,k,v} & \cdots & & & & & & & & & & & & \\
E_{1}^{k,v} & & \mathbf{0} & \ddots & & & & & & & & & & & & \\
\vdots & & & \ddots & \ddots & & & & & & & & & & & \\
E_{N-1}^{k,v} & & & & \mathbf{0} & \mathbf{D}_{N-1,N}^{Y,k,v} & & & & & & & & & & \\
E_{N}^{k,v} & & & & & \mathbf{0} & \mathbf{D}_{N,N+1}^{Y,k,v} & & & & & & & & & \\
E_{N+1}^{k,v} & & & & & & \mathbf{0} & \ddots & & & & & & & & \\
\vdots & & & & & & & \ddots & \ddots & & & & & & & \\
E_{k-1}^{k,v} & & & & & & & & \mathbf{0} & \mathbf{D}_{k-1,k}^{Y,k,v} & & & & & & \\
E_{k}^{k,v} & & & & & & & & & \mathbf{0} & & & & & & \\
\hline
E_{N}^{k,nv} & & & & & & & & & & & \mathbf{0} & \mathbf{D}_{N,N+1}^{Y,k,nv} & & & \\
E_{N+1}^{k,nv} & & & & & & & & & & & & \mathbf{0} & \mathbf{D}_{N+1,N+2}^{Y,k,nv} & & \\
\vdots & & & & & & & & & & & & & \ddots & \ddots & \\
E_{k-1}^{k,nv} & & & & & & & & & & & & & & \mathbf{0} & \mathbf{D}_{k-1,k}^{Y,k,nv}\\
E_{k}^{k,nv} & & & & & & & & & & & & & & & \mathbf{0}
\end{array}.
$$

These block matrices are for $k = 1,\ldots, n$ and $R > 1$,

$$
\mathbf{D}_{01}^{A,k,nv}=\left(\left(I_{\{k>1\}}\mathbf{H}_{A}+I_{\{k=1\}}\mathbf{H}'_{A}\right)\otimes\boldsymbol{\beta}_{1},\mathbf{0}\right) \quad ; \quad \mathbf{D}_{01}^{B,k,nv}=\left(\mathbf{0},\left(I_{\{k>1\}}\mathbf{H}_{B}+I_{\{k=1\}}\mathbf{H}'_{B}\right)\otimes\boldsymbol{\beta}_{2}\right).
$$

- For $r = \max\{1, k\!-\!R\!+\!1\}, \ldots, k\!-\!1$

$$\mathbf{D}_{r,r+1}^{A,k,nv} = \begin{pmatrix} \mathbf{CR}_{r,r+1}^{A,k,nv} \\ \mathbf{PM}_{r,r+1}^{A,k,nv} \end{pmatrix}, \ \mathbf{CR}_{r,r+1}^{A,k,nv} = \left( \mathbf{I}_{2^{r-1}} \otimes \left( \left( I_{\{r<k-1\}} \mathbf{H}_A + I_{\{r=k-1\}} \mathbf{H'}_A \right) \otimes \mathbf{I}, \mathbf{0} \right) \quad \mathbf{0} \right),$$

$$\mathbf{CR}_{r,r+1}^{A,k,nv} = \left( \mathbf{0} \quad \mathbf{I}_{2^{r-1}} \otimes \left( \left( I_{\{r<k-1\}} \mathbf{H}_A + I_{\{r=k-1\}} \mathbf{H'}_A \right) \otimes \mathbf{I}, \mathbf{0} \right) \right).$$

$$\mathbf{D}_{r,r+1}^{B,k,nv} = \begin{pmatrix} \mathbf{CR}_{r,r+1}^{B,k,nv} \\ \mathbf{PM}_{r,r+1}^{B,k,nv} \end{pmatrix}, \ \mathbf{CR}_{r,r+1}^{B,k,nv} = \left( \mathbf{I}_{2^{r-1}} \otimes \left( \mathbf{0}, \left( I_{\{r<k-1\}} \mathbf{H}_B + I_{\{r=k-1\}} \mathbf{H'}_B \right) \otimes \mathbf{I} \right) \quad \mathbf{0} \right),$$

$$\mathbf{PM}_{r,r+1}^{B,k,nv} = \left( \mathbf{0} \quad \mathbf{I}_{2^{r-1}} \otimes \left( \mathbf{0}, \left( I_{\{r<k-1\}} \mathbf{H}_B + I_{\{r=k-1\}} \mathbf{H'}_B \right) \otimes \mathbf{I} \right) \right).$$

- For $r = 1, \ldots, k\!-\!1$ and $k \geq R$

$$\mathbf{D}_{0,1}^{A,k,v} = \left( \mathbf{H}_A \otimes \mathbf{I}, \mathbf{0} \right); \ \mathbf{D}_{0,1}^{B,k,v} = \left( \mathbf{0}, \mathbf{H}_B \otimes \mathbf{I} \right), \ \mathbf{D}_{r,r+1}^{A,k,v} = \mathbf{I}_{2^r} \otimes \left( \left( I_{\{r<k-1\}} \mathbf{H}_A + I_{\{r=k-1\}} \mathbf{H'}_A \right) \otimes \mathbf{I}, \mathbf{0} \right),$$

$$\mathbf{D}_{r,r+1}^{B,k,v} = \mathbf{I}_{2^r} \otimes \left( \mathbf{0}, \left( I_{\{r<k-1\}} \mathbf{H}_B + I_{\{r=k-1\}} \mathbf{H'}_B \right) \otimes \mathbf{I} \right).$$

*The matrix* $\mathbf{D}^{D}$

$\mathbf{D}^{D}$: transition intensities for scenarios where the repairperson returns to work and stays at his workplace. The structure of this matrix is

$$\mathbf{D}^D = \begin{pmatrix} \mathbf{D}_n^D & & & & & & \\ & \mathbf{D}_{n-1}^D & & & & & \\ & & \ddots & & & & \\ & & & \mathbf{D}_R^D & & & \\ & & & & \mathbf{0} & & \\ & & & & & \ddots & \\ & & & & & & \mathbf{0} \end{pmatrix},$$

where for $k = R, \ldots, n$

$$
\mathbf{D}_k^D = \begin{array}{c} E_0^{k,v} \\ E_1^{k,v} \\ \vdots \\ E_{N-1}^{k,v} \\ E_N^{k,v} \\ E_{N+1}^{k,v} \\ \vdots \\ E_{k-1}^{k,v} \\ E_k^{k,v} \\ E_N^{k,nv} \\ E_{N+1}^{k,nv} \\ \vdots \\ E_{k-1}^{k,nv} \\ E_k^{k,nv} \end{array}
\overset{\begin{array}{cccccccccccccc} E_0^{k,v} & E_1^{k,v} & \cdots & E_{N-1}^{k,v} & E_N^{k,v} & E_{N+1}^{k,v} & \cdots & E_{k-1}^{k,v} & E_k^{k,v} & E_N^{k,nv} & E_{N+1}^{k,nv} & \cdots & E_{k-1}^{k,nv} & E_k^{k,nv} \end{array}}{\left(\begin{array}{c|cccc}
 & & & & \\
 & & & & \\
 & & & & \\
 & & & & \\
 & \mathbf{D}_{N,N}^{D,k,nv} & & & \\
 & & \mathbf{D}_{N+1,N+1}^{D,k,nv} & & \\
 & & & \ddots & \\
 & & & & \mathbf{D}_{k-1,k-1}^{D,k,nv} \\
 & & & & \qquad \mathbf{D}_{k,k}^{D,k,nv} \\
\hline
 & & & & \\
 & & & & \\
 & & & & \\
 & & & & \\
 & & & &
\end{array}\right)}.
$$

with

$$
\mathbf{D}_{r,r}^{D,k,nv} = \begin{pmatrix} \mathbf{CR}_{r,r}^{D,k,nv} \\ \mathbf{PM}_{r,r}^{D,k,nv} \end{pmatrix}, \ \mathbf{CR}_{r,r}^{D,k,nv} = \left( \mathbf{I}_{2^{r-1}} \otimes \left( I_{\{r<k\}} \mathbf{I}_{mtd\varepsilon} + I_{\{r=k\}} \mathbf{I}_t \right) \otimes \mathbf{V}^0 \otimes \boldsymbol{\beta}_1 \quad \mathbf{0} \right),
$$

$$
\mathbf{PM}_{r,r}^{D,k,nv} = \left( \mathbf{0} \quad \mathbf{I}_{2^{r-1}} \otimes \left( I_{\{r<k\}} \mathbf{I}_{mtd\varepsilon} + I_{\{r=k\}} \mathbf{I}_t \right) \otimes \mathbf{V}^0 \otimes \boldsymbol{\beta}_2 \right).
$$

*The matrix* $\mathbf{D}^{CD}$

$\mathbf{D}^{CD}$: transition intensities for scenarios where the repairperson has to interrupt the vacation due to a non-repairable failure. The structure of this matrix is

$$
\mathbf{D}^{CD} = \begin{pmatrix}
\mathbf{0} & \mathbf{0} & & & & \\
 & \mathbf{0} & \ddots & & & \\
 & & \mathbf{0} & \mathbf{D}_R^{CD} & & \\
 & & & \ddots & \ddots & \\
 & & & & \mathbf{0} & \mathbf{0} \\
\mathbf{0} & & & & \mathbf{0} & \mathbf{0}
\end{pmatrix},
$$

- for $k = R \geq 2$

$$\mathbf{D}_R^{CD} = \begin{array}{c} \\ E_0^{R,v} \\ E_1^{R,v} \\ \vdots \\ E_{R-1}^{R,v} \\ E_R^{R,v} \\ E_1^{R,nv} \\ E_2^{R,nv} \\ \vdots \\ E_{R-1}^{R,nv} \\ E_R^{R,nv} \end{array} \begin{array}{c} \begin{array}{ccccc} E_0^{R-1,nv} & E_1^{R-1,nv} & \cdots & E_{R-2}^{R-1,nv} & E_{R-1}^{R-1,nv} \end{array} \\ \left( \begin{array}{ccccc} \mathbf{D}_{00}^{CD,R,nv} & & & & \mathbf{0} \\ & \mathbf{D}_{11}^{CD,R,nv} & & & \\ & & \ddots & & \\ & & & & \mathbf{D}_{R-1,R-1}^{CD,R,nv} \\ \mathbf{0} & & & & \mathbf{0} \\ \hline \mathbf{0} & \mathbf{0} & & & \\ & \mathbf{0} & \ddots & & \\ & & \ddots & \ddots & \\ & & & \mathbf{0} & \mathbf{0} \\ \mathbf{0} & \cdots & & \cdots & \mathbf{0} \end{array} \right) \end{array},$$

$$\mathbf{D}_{00}^{CD,R,nv} = \mathbf{H}_C \otimes \mathbf{e}_v.$$

- for $r = 1, \ldots, R-1$,

$$\mathbf{D}_{r,r}^{CD,R,nv} = \begin{pmatrix} \mathbf{CR}_{r,r}^{CD,R,nv} \\ \mathbf{PM}_{r,r}^{CD,R,nv} \end{pmatrix}, \ \mathbf{CR}_{r,r}^{CD,R,nv} = \left( \mathbf{I}_{2^{r-1}} \otimes \left( I_{\{r=R-1\}} \mathbf{H}'_C + I_{\{r<R-1\}} \mathbf{H}_C \right) \otimes \mathbf{e}_v \otimes \boldsymbol{\beta}_1 \quad \mathbf{0} \right),$$

$$\mathbf{PM}_{r,r}^{CD,R,nv} = \left( \mathbf{0} \quad \mathbf{I}_{2^{r-1}} \otimes \left( I_{\{r=R-1\}} \mathbf{H}'_C + I_{\{r<R-1\}} \mathbf{H}_C \right) \otimes \mathbf{e}_v \otimes \boldsymbol{\beta}_2 \right).$$

*The matrix* $\mathbf{D}^E$

$\mathbf{D}^E$: transition intensities for the case, the repairperson returns to work and begins a new vacation period. The matrix is

$$\mathbf{D}^E = \begin{pmatrix} \mathbf{D}_n^E & & & & & & \\ & \mathbf{D}_{n-1}^E & & & & & \\ & & \ddots & & & & \\ & & & \mathbf{D}_R^E & & & \\ & & & & \mathbf{0} & & \\ & & & & & \ddots & \\ & & & & & & \mathbf{0} \end{pmatrix}, \text{ with}$$

$$
\mathbf{D}_k^E = \begin{array}{c} \\ E_0^{k,v} \\ E_1^{k,v} \\ \vdots \\ E_{N-1}^{k,v} \\ E_N^{k,v} \\ E_{N+1}^{k,v} \\ \vdots \\ E_{k-1}^{k,v} \\ E_k^{k,v} \\ E_N^{k,nv} \\ E_{N+1}^{k,nv} \\ \vdots \\ E_{k-1}^{k,nv} \\ E_k^{k,nv} \end{array}
\begin{array}{c}
\begin{array}{cccccccccc:ccccc} E_0^{k,v} & E_1^{k,v} & \cdots & E_{N-1}^{k,v} & E_N^{k,v} & E_{N+1}^{k,v} & \cdots & E_{k-1}^{k,v} & E_k^{k,v} & & E_N^{k,nv} & E_{N+1}^{k,nv} & \cdots & E_{k-1}^{k,nv} & E_k^{k,nv} \end{array} \\
\left(\begin{array}{ccccccccc:ccccc}
\mathbf{D}_{0,0}^{E,k,v} & & & & & & & & & & & & & \\
 & \mathbf{D}_{1,1}^{E,k,v} & & & & & & & & & & & & \\
 & & \ddots & & & & & & & & & & & \\
 & & & \mathbf{D}_{N-1,N-1}^{E,k,v} & & & & & & & & & & \\
 & & & & & & & & & & & & & \\
 & & & & & & & & & & & & & \\
 & & & & & & & & & & & & & \\
 & & & & & & & & & & & & & \\
 & & & & & & & & & & & & & \\
\hdashline
 & & & & & & & & & & & & & \\
 & & & & & & & & & & & & & \\
 & & & & & & & & & & & & & \\
 & & & & & & & & & & & & & \\
 & & & & & & & & & & & & &
\end{array}\right)
\end{array},
$$

$$
\mathbf{D}_{0,0}^{E,k,v} = \mathbf{I} \otimes \mathbf{V}^0 \boldsymbol{\upsilon} \text{ and } \mathbf{D}_{r,r}^{E,k,v} = \begin{pmatrix} \mathbf{CR}_{r,r}^{E,k,v} \\ \mathbf{PM}_{r,r}^{E,k,v} \end{pmatrix}, \text{ where}
$$

$$
\mathbf{CR}_{r,r}^{E,k,v} = \begin{pmatrix} \mathbf{I}_{2^{r-1}} \otimes \mathbf{I} \otimes \mathbf{V}^0 \boldsymbol{\upsilon} & \mathbf{0} \end{pmatrix} \text{ and } \mathbf{PM}_{r,r}^{E,k,v} = \begin{pmatrix} \mathbf{0} & \mathbf{I}_{2^{r-1}} \otimes \mathbf{I} \otimes \mathbf{V}^0 \boldsymbol{\upsilon} \end{pmatrix}, \text{ for } r = 1, \ldots, N-1.
$$

*The matrix* $\mathbf{D}^F$

$\mathbf{D}^F$: transition intensities for the case: repairperson finishes a repair and begins a new vacation period. The structure is

$$
\mathbf{D}^F = \begin{pmatrix}
\mathbf{D}_n^F & & & & & & \\
 & \mathbf{D}_{n-1}^F & & & & & \\
 & & \ddots & & & & \\
 & & & \mathbf{D}_R^F & & & \\
 & & & & \mathbf{0} & & \\
 & & & & & \ddots & \\
 & & & & & & \mathbf{0}
\end{pmatrix},
$$

with

$$
\mathbf{D}_k^F = \begin{array}{c} \\ E_0^{k,v} \\ E_1^{k,v} \\ \vdots \\ E_{N-1}^{k,v} \\ E_N^{k,v} \\ E_{N+1}^{k,v} \\ \vdots \\ E_{k-1}^{k,v} \\ E_k^{k,v} \\ E_N^{k,nv} \\ E_{N+1}^{k,nv} \\ \vdots \\ E_{k-1}^{k,nv} \\ E_k^{k,nv} \end{array}
\begin{array}{c}
\begin{array}{cccccccccccccc} E_0^{k,v} & E_1^{k,v} & \dots & E_{N-1}^{k,v} & E_N^{k,v} & E_{N+1}^{k,v} & \dots & E_{k-1}^{k,v} & E_k^{k,v} & E_N^{k,nv} & E_{N+1}^{k,nv} & \dots & E_{k-1}^{k,nv} & E_k^{k,nv} \end{array} \\
\left(\begin{array}{ccccccccc:ccccc}
 & & & & & & & & & & & & & \\
 & & & & & & & & & & & & & \\
 & & & & & & & & & & & & & \\
 & & & & & & & & & & & & & \\
 & & & & & & & & & & & & & \\
 & & & & & & & & & & & & & \\
 & & & & & & & & & & & & & \\
 & & & & & & & & & & & & & \\
 & & & & & & & & & & & & & \\
\hdashline
 & & & \mathbf{D}_{N,N-1}^{F,k,v} & & & & & & & & & & \\
 & & & & & & & & & & & & & \\
 & & & & & & & & & & & & & \\
 & & & & & & & & & & & & & \\
 & & & & & & & & & & & & &
\end{array}\right)
\end{array},
$$

$$
\mathbf{D}_{N,N-1}^{F,k,v} = \begin{pmatrix} \mathbf{I}_{2^{N-1}} \otimes \left( I_{\{k=N\}} \boldsymbol{\theta} + I_{\{k>N\}} \mathbf{I} \right) \otimes \boldsymbol{\upsilon} \otimes \mathbf{S}_1^0 \\ \mathbf{I}_{2^{N-1}} \otimes \left( I_{\{k=N\}} \boldsymbol{\theta} + I_{\{k>N\}} \mathbf{I} \right) \otimes \boldsymbol{\upsilon} \otimes \mathbf{S}_2^0 \end{pmatrix}.
$$

*The matrix* $\mathbf{D}^{NS}$

$\mathbf{D}^{NS}$: transition intensities when a failure provokes a new system. It is

$$
\mathbf{D}^{NS} = \begin{pmatrix} \mathbf{0} & & & & \\ & \mathbf{0} & & & \\ & & \mathbf{0} & & \\ & & & \ddots & \\ \mathbf{D}_1^{NS} & & & & \mathbf{0} \end{pmatrix}, \text{ where}
$$

- if $R = 1$

$$
\mathbf{D}_1^{NS} = \begin{array}{c} E_0^{1,v} \\ E_1^{1,v} \\ E_1^{1,nv} \end{array}
\overset{\begin{array}{cccccccc} E_0^{n,v} & E_1^{n,v} & \dots & E_{N-1}^{n,v} & E_N^{n,v} & E_{N+1}^{n,v} & \dots & E_n^{n,v} \; E_n^{n,nv} \end{array}}{\begin{pmatrix} \mathbf{D}_{00}^{NS,1,v} & \mathbf{0} & \dots & & \mathbf{0} \\ \mathbf{0} & \mathbf{0} & \dots & & \mathbf{0} \\ \mathbf{0} & \mathbf{0} & \dots & & \mathbf{0} \end{pmatrix}},
$$

with $\mathbf{D}_{00}^{NS,1,v} = \mathbf{H}_C \otimes \mathbf{e}_\upsilon \mathbf{v}$.

- If $R > 1$

$$
\mathbf{D}_1^{NS} = \begin{array}{c} E_0^{1,nv} \\ E_1^{1,nv} \end{array}
\overset{\begin{array}{cccccccccccc} E_0^{n,v} & E_1^{n,v} & \dots & E_{N-1}^{n,v} & E_N^{n,v} & E_{N+1}^{n,v} & \dots & E_k^{n,v} & E_{N=n-R+1}^{n,nv} & E_{N+1}^{n,nv} & \dots & E_{n-1}^{n,nv} \; E_n^{n,nv} \end{array}}{\begin{pmatrix} \mathbf{D}_{00}^{NS,1,v} & \mathbf{0} & \dots & & \dots & \mathbf{0} \\ \mathbf{0} & \mathbf{0} & \dots & & \dots & \mathbf{0} \end{pmatrix}},
$$

with $\mathbf{D}_{00}^{NS,1,v} = \mathbf{H}_C \otimes \mathbf{v}$.

*The matrix* $\mathbf{D}^O$

$\mathbf{D}^O$: transition intensities when none event occurs (given the MMAP).

$$
\mathbf{D}^O = \begin{pmatrix} \mathbf{D}_n^O & & & & \\ & \mathbf{D}_{n-1}^O & & & \\ & & \mathbf{D}_{n-2}^O & & \\ & & & \ddots & \\ & & & & \mathbf{D}_1^O \end{pmatrix},
$$

- For $k = 1,\dots, R-1$

$$
\mathbf{D}_k^O = \begin{array}{c} \\ E_0^{k,nv} \\ E_1^{k,nv} \\ \vdots \\ E_{k-1}^{k,vn} \\ E_k^{k,nv} \end{array}
\begin{array}{c}
\begin{array}{ccccc} E_0^{k,nv} & E_1^{k,nv} & E_2^{k,nv} & E_{k-1}^{k,nv} & E_k^{k,nv} \end{array} \\
\begin{pmatrix}
\mathbf{D}_{00}^{O,k,nv} & & & & \\
\mathbf{D}_{10}^{O,k,nv} & \mathbf{D}_{11}^{O,k,nv} & & & \\
 & \ddots & \ddots & & \\
 & & \mathbf{D}_{k-1,k-2}^{O,k,nv} & \mathbf{D}_{k-1,k-1}^{O,k,nv} & \\
 & & & \mathbf{D}_{k,k-1}^{O,k,nv} & \mathbf{D}_{k,k}^{O,k,nv}
\end{pmatrix}
\end{array}.
$$

- For $k = R, \ldots, n$

$$
\mathbf{D}_k^O = \begin{array}{c} \\ E_0^{k,v} \\ E_1^{k,v} \\ \vdots \\ E_{N-1}^{k,v} \\ E_N^{k,v} \\ E_{N+1}^{k,v} \\ \vdots \\ E_{k-1}^{k,v} \\ E_k^{k,v} \\ E_N^{k,nv} \\ E_{N+1}^{k,nv} \\ \vdots \\ E_{k-1}^{k,nv} \\ E_k^{k,nv} \end{array}
\begin{array}{c}
\begin{array}{ccccccccccccc} E_0^{k,v} & E_1^{k,v} & \ldots & E_{N-1}^{k,v} & E_N^{k,v} & E_{N+1}^{k,v} & \ldots & E_{k-1}^{k,v} & E_k^{k,v} & E_N^{k,nv} & E_{N+1}^{k,nv} & E_{N+2}^{k,nv} & E_{k-1}^{k,nv} & E_k^{k,nv} \end{array} \\
\left(\begin{array}{ccccccccc|ccccc}
\mathbf{D}_{00}^{O,k,v} & & \ldots & & & & & & & & & & & \\
 & \mathbf{D}_{11}^{O,k,v} & & & & & & & & & & & & \\
 & & \ddots & & & & & & & & & & & \\
 & & & \mathbf{D}_{N-1,N-1}^{O,k,v} & & & & & & & & & & \\
 & & & & \mathbf{D}_{N,N}^{O,k,v} & & & & & & & & & \\
 & & & & & \mathbf{D}_{N+1,N+1}^{O,k,v} & & & & & & & & \\
 & & & & & & \ddots & & & & & & & \\
 & & & & & & & \mathbf{D}_{k-1,k-1}^{O,k,v} & & & & & & \\
 & & & & & & & & \mathbf{D}_{k,k}^{O,k,v} & & & & & \\
\hline
 & & & & & & & & & \mathbf{D}_{N,N}^{O,k,nv} & & & & \\
 & & & & & & & & & \mathbf{D}_{N+1,N}^{O,k,nv} & \mathbf{D}_{N+1,N+1}^{O,k,nv} & & & \\
 & & & & & & & & & & \ddots & \ddots & & \\
 & & & & & & & & & & & \mathbf{D}_{k-1,k-2}^{O,k,nv} & \mathbf{D}_{k-1,k-1}^{O,k,nv} & \\
 & & & & & & & & & & & & \mathbf{D}_{k,k-1}^{O,k,nv} & \mathbf{D}_{k,k}^{O,k,nv}
\end{array}\right)
\end{array},
$$

with $\boldsymbol{\theta} = \boldsymbol{\alpha} \otimes \mathbf{I} \otimes \boldsymbol{\omega} \otimes \boldsymbol{\eta}$.

- For $r = 0, \ldots, k$

$$
\mathbf{D}_{r,r}^{O,k,v} = \mathbf{I}_{2^r} \otimes \left( I_{\{r<k\}} \mathbf{H}_O + I_{\{r=k\}} \left( \mathbf{L} + \mathbf{L}^0 \boldsymbol{\gamma} \right) \right) \oplus \mathbf{V},
$$

$$
\mathbf{D}_{00}^{O,k,nv} = \mathbf{H}_O.
$$

- For $r = 1, \ldots, k$

$$
\mathbf{D}_{r,r}^{O,k,nv} = \begin{pmatrix} \mathbf{CR}_{r,r}^{O,k,nv} \\ \mathbf{PM}_{r,r}^{O,k,nv} \end{pmatrix}, \quad \mathbf{CR}_{r,r}^{O,k,nv} = \left( \mathbf{I}_{2^{r-1}} \otimes \left( I_{\{r<k\}} \mathbf{H}_O + I_{\{r=k\}} \left( \mathbf{L} + \mathbf{L}^0 \boldsymbol{\gamma} \right) \right) \oplus \mathbf{S}_1 \quad \mathbf{0} \right),
$$

$$
\mathbf{PM}_{r,r}^{O,k,nv} = \left( \mathbf{0} \quad \mathbf{I}_{2^{r-1}} \otimes \left( I_{\{r<k\}} \mathbf{H}_O + I_{\{r=k\}} \left( \mathbf{L} + \mathbf{L}^0 \boldsymbol{\gamma} \right) \right) \oplus \mathbf{S}_2 \right).
$$

$$\mathbf{D}_{10}^{O,k,nv}=\begin{pmatrix}\mathbf{CR}_{10}^{O,k,nv}\\ \mathbf{PM}_{10}^{O,k,nv}\end{pmatrix},\ \mathbf{CR}_{10}^{O,k,nv}=\left(I_{\{k>1\}}\mathbf{I}+I_{\{k=1\}}\boldsymbol{\theta}\right)\otimes\mathbf{S}_1^0,\ \mathbf{PM}_{10}^{O,k,nv}=\left(I_{\{k>1\}}\mathbf{I}+I_{\{k=1\}}\boldsymbol{\theta}\right)\otimes\mathbf{S}_2^0.$$

- For $r = 2,\dots, k$

$$\mathbf{D}_{r,r-1}^{O,k,nv}=\begin{pmatrix}\mathbf{CRCR}_{r,r-1}^{O,k,nv}\\ \mathbf{CRPM}_{r,r-1}^{O,k,nv}\\ \mathbf{PMCR}_{r,r-1}^{O,k,nv}\\ \mathbf{PMPM}_{r,r-1}^{O,k,nv}\end{pmatrix},$$

$$\mathbf{CRCR}_{r,r-1}^{O,k,nv}=\left(I_{2^{r-2}}\otimes\left(I_{\{r<k\}}\mathbf{I}+I_{\{r=k\}}\boldsymbol{\theta}\right)\otimes\mathbf{S}_1^0\otimes\boldsymbol{\beta}_1\quad \mathbf{0}\right),$$

$$\mathbf{CRPM}_{r,r-1}^{O,k,nv}=\left(\mathbf{0}\quad I_{2^{r-2}}\otimes\left(I_{\{r<k\}}\mathbf{I}+I_{\{r=k\}}\boldsymbol{\theta}\right)\otimes\mathbf{S}_1^0\otimes\boldsymbol{\beta}_2\right),$$

$$\mathbf{PMCR}_{r,r-1}^{O,k,nv}=\left(I_{2^{r-2}}\otimes\left(I_{\{r<k\}}\mathbf{I}+I_{\{r=k\}}\boldsymbol{\theta}\right)\otimes\mathbf{S}_2^0\otimes\boldsymbol{\beta}_1\quad \mathbf{0}\right),$$

$$\mathbf{PMPM}_{r,r-1}^{O,k,nv}=\left(\mathbf{0}\quad I_{2^{r-2}}\otimes\left(I_{\{r<k\}}\mathbf{I}+I_{\{r=k\}}\boldsymbol{\theta}\right)\otimes\mathbf{S}_2^0\otimes\boldsymbol{\beta}_2\right).$$

*Appendix C*

The following example illustrates the methodology for the profit/cost vector developed in Section 4.4.1.

**Example 1**. Consider a system with $m = d = v = 2$ and $t = \varepsilon = 1$. Therefore, there are two internal operational phases, with an associated cost of 1 for phase 1 and 2 for phase 2, i.e., $\mathbf{c}_0=(1,2)'$. Additionally, we assume that there are two wear phases associated with external shocks, where the cost incurred during these phases is given by $\mathbf{c}_d=(0,1)'$. In this example, it is also assumed that corrective repair involves two stages ($z_1$=2) with costs $\mathbf{cr}_1=(0.2,0.5)'$. Preventive maintenance involves a single stage ($z_1$=2) with a cost $\mathbf{cr}_2=0.1$.

In this example, we will construct some specific cases of the cost vectors associated with macro-states, directly illustrating the algorithmic process.

First, we develop the case $\mathbf{nr}_0^{k,v}$ associated with macro-state $E_0^{k,v}$. The phases associated with this macro-state are defined as $(k,0;i,j,h,u,w)$ con $i$=1, 2; $j$=1 ; $h$=1,2; $u$=1 and $w$=1, 2 resulting in

$$\mathbf{E}_0^{k,v} = \begin{matrix}(k,0;1,1,1,1,1)\\(k,0;1,1,1,1,2)\\(k,0;1,1,2,1,1)\\(k,0;1,1,2,1,2)\\(k,0;2,1,1,1,1)\\(k,0;2,1,1,1,2)\\(k,0;2,1,2,1,1)\\(k,0;2,1,2,1,2)\end{matrix}.$$

Thus, the net profits for these phases are given by

$$\mathbf{nr}_0^{k,v} = \begin{pmatrix}B-F-1-0\\B-F-1-0\\B-F-1-1\\B-F-1-1\\B-F-2-0\\B-F-2-0\\B-F-2-1\\B-F-2-1\end{pmatrix} = \begin{pmatrix}B-F-1\\B-F-1\\B-F-2\\B-F-2\\B-F-2\\B-F-2\\B-F-3\\B-F-3\end{pmatrix} = (B-F)\mathbf{e}_8 - \begin{pmatrix}1\\2\end{pmatrix}\otimes\mathbf{e}_4 - \mathbf{e}_2\otimes\begin{pmatrix}0\\1\end{pmatrix}\otimes\mathbf{e}_2.$$

What happens when there are units in the repair channel? Consider the case *s*=2, where there are two units in the repair channel. Clearly, the level 2 macro-state is expressed in terms of the level 3 macro-states as $\mathbf{E}_2^{k,v} = \begin{pmatrix}\mathbf{E}_{1,1}^{k,v}\\\mathbf{E}_{1,2}^{k,v}\\\mathbf{E}_{2,1}^{k,v}\\\mathbf{E}_{2,2}^{k,v}\end{pmatrix}$. The phases for these level 3 macro-states are $(k,2;i,j,h,u,w)$ con *i*=1, 2; *j*=1 ; *h*=1,2; *u*=1 and *w*=1,2. Therefore,

$$\mathbf{E}_{1,1}^{k,v} = \mathbf{E}_{1,2}^{k,v} = \mathbf{E}_{2,1}^{k,v} = \mathbf{E}_{2,2}^{k,v} = \begin{matrix}(k,2;1,1,1,1,1)\\(k,2;1,1,1,1,2)\\(k,2;1,1,2,1,1)\\(k,2;1,1,2,1,2)\\(k,2;2,1,1,1,1)\\(k,2;2,1,1,1,2)\\(k,2;2,1,2,1,1)\\(k,2;2,1,2,1,2)\end{matrix}.$$

Thus, the net benefits for these phases are

$$\mathbf{nr}_{1,1}^{k,v}=\mathbf{nr}_{1,2}^{k,v}=\mathbf{nr}_{2,1}^{k,v}=\mathbf{nr}_{2,1}^{k,v}=\begin{pmatrix} B-F-1-0 \\ B-F-1-0 \\ B-F-1-1 \\ B-F-1-1 \\ B-F-2-0 \\ B-F-2-0 \\ B-F-2-1 \\ B-F-2-1 \end{pmatrix}=\begin{pmatrix} B-F-1 \\ B-F-1 \\ B-F-2 \\ B-F-2 \\ B-F-2 \\ B-F-2 \\ B-F-3 \\ B-F-3 \end{pmatrix}$$

$$=(B-F)\mathbf{e}_8-\begin{pmatrix}1\\2\end{pmatrix}\otimes\mathbf{e}_4-\mathbf{e}_2\otimes\begin{pmatrix}0\\1\end{pmatrix}\otimes\mathbf{e}_2$$

$$=(B-F)\mathbf{e}_8-\mathbf{c}_0\otimes\mathbf{e}_4-\mathbf{e}_2\otimes\mathbf{c}_d\otimes\mathbf{e}_2$$

Consequently, $\mathbf{nr}_2^{k,v}=\begin{pmatrix}\mathbf{nr}_{1,1}^{k,v}\\ \mathbf{nr}_{1,2}^{k,v}\\ \mathbf{nr}_{2,1}^{k,v}\\ \mathbf{nr}_{2,2}^{k,v}\end{pmatrix}=\mathbf{e}_4\otimes\left((B-F)\mathbf{e}_8-\mathbf{c}_0\otimes\mathbf{e}_4-\mathbf{e}_2\otimes\mathbf{c}_d\otimes\mathbf{e}_2\right).$

Next, let's consider a case where the repairer is in the repair channel with $s$=3, meaning three units are in the repair facility ($k$>3). The level 2 macro-state, $\mathbf{E}_3^{k,nv}$, is composed of the level 3 macro-states $\mathbf{E}_{i_1,i_2,i_3}^{k,nv}$ as follows,

$$\mathbf{E}_3^{k,nv}=\begin{matrix}\mathbf{E}_{1,1,1}^{k,nv}\\ \mathbf{E}_{1,1,2}^{k,nv}\\ \mathbf{E}_{1,2,1}^{k,nv}\\ \mathbf{E}_{1,2,2}^{k,nv}\\ \mathbf{E}_{2,1,1}^{k,nv}\\ \mathbf{E}_{2,1,2}^{k,nv}\\ \mathbf{E}_{2,2,1}^{k,nv}\\ \mathbf{E}_{2,2,2}^{k,nv}\end{matrix}.$$

The phases of these level 3 macro-states are given by $(k,3;i,j,h,u,r)$ with $i$=1, 2; $j$=1; $h$=1, 2; $\varepsilon$=1; and $r$=1, 2 if $i_1$=1 (corrective repair with two phases) or $r$=1 if $i_1$=2 (preventice maintenance with one phase). Therefore,

$$\mathbf{E}_{1,1,1}^{k,nv} = \mathbf{E}_{1,1,2}^{k,nv} = \mathbf{E}_{1,2,1}^{k,nv} = \mathbf{E}_{1,2,2}^{k,nv} = \begin{matrix} k,3;1,1,1,1,1 \\ k,3;1,1,1,1,2 \\ k,3;1,1,2,1,1 \\ k,3;1,1,2,1,2 \\ k,3;2,1,1,1,1 \\ k,3;2,1,1,1,2 \\ k,3;2,1,2,1,1 \\ k,3;2,1,2,1,2 \end{matrix},$$

$$\mathbf{E}_{2,1,1}^{k,nv} = \mathbf{E}_{2,1,2}^{k,nv} = \mathbf{E}_{2,2,1}^{k,nv} = \mathbf{E}_{2,2,2}^{k,nv} = \begin{matrix} k,3;1,1,1,1,1 \\ k,3;1,1,2,1,1 \\ k,3;2,1,1,1,1 \\ k,3;2,1,2,1,1 \end{matrix}.$$

Thus, the net benefits for these phases are

$$\mathbf{nc}_{1,1,1}^{k,nv} = \mathbf{nc}_{1,1,2}^{k,nv} = \mathbf{nc}_{1,2,1}^{k,nv} = \mathbf{nc}_{1,2,2}^{k,nv} = \begin{pmatrix} 0.2 \\ 0.5 \\ 0.2 \\ 0.5 \\ 0.2 \\ 0.5 \\ 0.2 \\ 0.5 \end{pmatrix} = \mathbf{e}_4 \otimes \begin{pmatrix} 0.2 \\ 0.5 \end{pmatrix} = \mathbf{e}_4 \otimes \mathbf{cr}_1,$$

$$\mathbf{nc}_{2,1,1}^{k,nv} = \mathbf{nc}_{2,1,2}^{k,nv} = \mathbf{nc}_{2,2,1}^{k,nv} = \mathbf{nc}_{2,2,2}^{k,nv} = \begin{pmatrix} 0.1 \\ 0.1 \\ 0.1 \\ 0.1 \end{pmatrix} = \mathbf{e}_4 \otimes 0.1 = \mathbf{e}_4 \otimes \mathbf{cr}_2.$$

Therefore,

$$\mathbf{nc}_3^{k,nv} = \begin{pmatrix} \mathbf{nc}_{1,1,1}^{k,nv} \\ \mathbf{nc}_{1,1,2}^{k,nv} \\ \mathbf{nc}_{1,2,1}^{k,nv} \\ \mathbf{nc}_{1,2,2}^{k,nv} \\ \mathbf{nc}_{2,1,1}^{k,nv} \\ \mathbf{nc}_{2,1,2}^{k,nv} \\ \mathbf{nc}_{2,2,1}^{k,nv} \\ \mathbf{nc}_{2,2,2}^{k,nv} \end{pmatrix} = \begin{pmatrix} \mathbf{e}_4 \otimes \mathbf{e}_4 \otimes \mathbf{cr}_1 \\ \mathbf{e}_4 \otimes \mathbf{e}_4 \otimes \mathbf{cr}_2 \end{pmatrix}.$$

In Section 4.4.2, the mean net profit up to time $t$ generated by the main unit, $\Phi_w(t)$, equation (6), and the mean costs up to time $t$ incurred in the repair channel, $\Phi_{rf}(t)$, equation (7), are calculated. In these functions, the vector $\int_0^t \mathbf{p}(t)\,dt$ is considered, with $\mathbf{p}(t)$ being the transient distribution vector of the system up to time $t$. Each element of this measure, the integral vector, represents the average time spent in the corresponding phase of the system during time $t$. By performing the dot product, the proposed measures are obtained. Equations (8) and (9) follow directly from the description given in the work.